\documentclass[%
reprint,
twocolumns,
aps,
pre,
amsmath,
amssymb,
floatfix,
showpacs,
notitlepage,
nobibnotes,
superscriptaddress]
{revtex4-2}
\usepackage{graphicx} 
\usepackage[colorlinks=true,linkcolor=blue,citecolor=blue,filecolor=cyan,urlcolor=blue,breaklinks=true]{hyperref}
\usepackage{mathtools}
\usepackage{amsmath}
\usepackage[table]{xcolor}
\usepackage{makecell}
\usepackage{array}
\usepackage{etoolbox}
\usepackage[normalem]{ulem}

\newcommand{\mean}[1]{\left\langle #1 \right\rangle}
\newcommand{\pstat}{p^\mathrm{s}}
\newcommand{\stat}{^\mathrm{s}}
\newcommand{\dN}{\langle\dot N\rangle}
\newcommand{\artanh}{\operatorname{artanh}}
\newcommand{\traf}{u}

\begin{document}

\title{Thermodynamic cost for precision of general counting observables}

\author{Patrick Pietzonka
}
\affiliation{Max Planck Institute for the Physics of Complex Systems, N\"othnitzer Stra\ss e 38, 01187 Dresden, Germany}
\affiliation{SUPA, School of Physics and Astronomy, University of Edinburgh, Peter Guthrie Tait Road, Edinburgh EH9 3FD, United Kingdom}

\author{Francesco Coghi
}
\affiliation{Nordita, KTH Royal Institute of Technology and Stockholm University, Hannes Alfvéns väg 12, SE-106 91 Stockholm, Sweden}

\date{\today}

\begin{abstract}
    
We analytically derive universal bounds that describe the trade-off between thermodynamic cost and precision in a sequence of events related to some internal changes of an otherwise hidden physical system. The precision is quantified by the fluctuations in either the number of events counted over time or the waiting times between successive events. Our results are valid for the same broad class of nonequilibrium driven systems considered by the thermodynamic uncertainty relation, but they extend to both time-symmetric and asymmetric observables. We show how optimal precision saturating the bounds can be achieved. For waiting time fluctuations of asymmetric observables, a phase transition in the optimal configuration arises, where higher precision can be achieved by combining several signals.

\end{abstract}

\maketitle

\section{Introduction}


An out-of-equilibrium system is kept in a black box and observed from the outside. The only accessible information about the system is given by a time series of discrete events, which we will refer to as ``ticks'' throughout the paper. These could be audible ticks of a hidden clock, changes of conformation of a biomolecule~\cite{schw97,qian04,stig11}, a spike train produced by a neuron~\cite{Bialek1991,Bialek1993,Bialek1996,Dayan2001,Stein2005,Faisal2008,Rajdl2020,Hasegawa2022}, or steps taken by a molecular motor~\cite{noji97,chen02,lipo07}. In all cases, the question we face is the same: how much does it cost, in terms of energy dissipated by the system, to generate a precise sequence of ticks?


Being precise despite the presence of thermal fluctuations comes at a minimal thermodynamic cost quantified by the rate of entropy production. This is the gist of the thermodynamic uncertainty relation (TUR)~\cite{bara15,ging16,horo19}. It applies to a vast class of systems driven into a non-equilibrium steady state, namely all systems that
can, at some microscopic scale, be modelled as a Markov jump process or as an
overdamped Brownian motion. However, the concept of precision in the original formulation of the TUR only applies to so-called integrated current observables $X(t)$, which are additive in time $t$ and anti-symmetric under time reversal, e.g., the distance travelled by a molecular motor or the number of molecules produced in a chemical reaction.

In our general setting, however, each tick is fully characterised only by the moment in time it has appeared. The full information we have on the system is the number of ticks $N(t)$ recorded up to time $t$. Since $N(t)$ counts the number of events, it is always positive, even under time reversal of the system dynamics. Therefore the TUR does not provide us with a tool to quantify the trade-off between precision and dissipation. As one of our main results, we derive a useful analogue of the TUR. This extends the quantification of the thermodynamic cost of precision from currents to time-symmetric and asymmetric counting observables.

The interplay between thermodynamic cost and various desirable or measurable properties of a non-equilibrium system has recently been considered in stochastic thermodynamics~\cite{seif12,peli21}: From the study of the performance in quickly changing a statistical distribution~\cite{ito20,vo20,nich20,delv21,fala22,van23}, to the minimisation of
fluctuations in first passage times~\cite{ging17,neri20,pal21,Neri2022}, and to inferring properties of a non-equilibrium system from partial measurements~\cite{aman10,alem15,pole17,gnes18,seif19,busi19,ehri21,rold21,Baiesi2023} or under coarse-graining~\cite{rold10,espo12,Teza2020,Bilotto2021,Ghosal2022,Ghosal2023}. We expand on this theme, deriving thermodynamic bounds on precision in a sequence of discrete events, generally characterised through the variance of the total count, the variance of the waiting time between events, or through large deviation functions~\cite{touc09}.

The information about a non-equilibrium system that feeds into our bounds is arguably minimal. We only consider the precision of ticks---an observable related to time-symmetric, or ``frenetic'' aspects~\cite{maes08,maes20}---regardless of how exactly they are produced. And yet, we find a relation to the overall entropy production, a time-antisymmetric quantity with a clear thermodynamic interpretation. Previously, such minimal observations could only be related to another frenetic quantity, namely the dynamical activity, which is the rate of \textit{all} microscopic transitions~\cite{garr17,terl18,hiur21}. Other bounds that do involve the rate of entropy production require more information, such as the covariance of a time-symmetric quantity with a current~\cite{dech21}, a current along with the dynamical activity \cite{vo22}, or in a run-and-tumble process a current along with run-length statistics~\cite{Shreshtha2019}. Recent efforts to infer entropy production from waiting-time data available for a few observable transitions~\cite{mart19,harv20,skin21,skin21a,meer22,haru22,Harunari2023} have led to strong bounds, but require more detailed information on the microscopic state (or class of states) before or after a transition. With the inequalities derived in this manuscript, an external observer can infer bounds on the entropy that are typically looser, but require significantly less detailed information. While our bounds may be of limited use as inference tools for generic counting observables in random Markov networks, they do become useful for biological systems or technological applications where the production of a reliable, time-periodic signal is crucial. By showing how our bounds can be saturated, we reveal the design principles for systems that do so optimally at a given energy budget. 

The paper is organised as follows. In Sec.~\ref{sec:setup}, we give a minimal description of the setup that allows us to state the main results in Sec.~\ref{sec:mainres}. We then show applications of these results to experimental data for neural spikes and atomic clocks in Sec.~\ref{sec:applications}. In Sec.~\ref{sec:illustration}, we illustrate our results for examples of Markov networks, show networks with optimal precision, and compare our results with the relevant literature. In the following sections, we focus on the case of ticks associated with a single edge, which already yields the correct mathematical form of all bounds (with one notable exception). Bounds on large deviation functions are derived in Sec.~\ref{sec:ldf}, before focusing on typical fluctuations in Sec.~\ref{sec:typ}.  In Sec.~\ref{sec:wtd} we derive a variational principle for waiting time distributions, which then yields analogous thermodynamic bounds on precision. In Sec.~\ref{sec:optimal}, we calculate the cost and precision exactly for a class of networks in which bounds can be saturated. We finally generalise all our results to the case of ticks stemming from several edges in Sec.~\ref{sec:multiedge}, and conclude in Sec.~\ref{sec:conclusions}. 

\section{Setup}
\label{sec:setup}

\begin{figure}
    \centering
    \includegraphics[width=\linewidth]{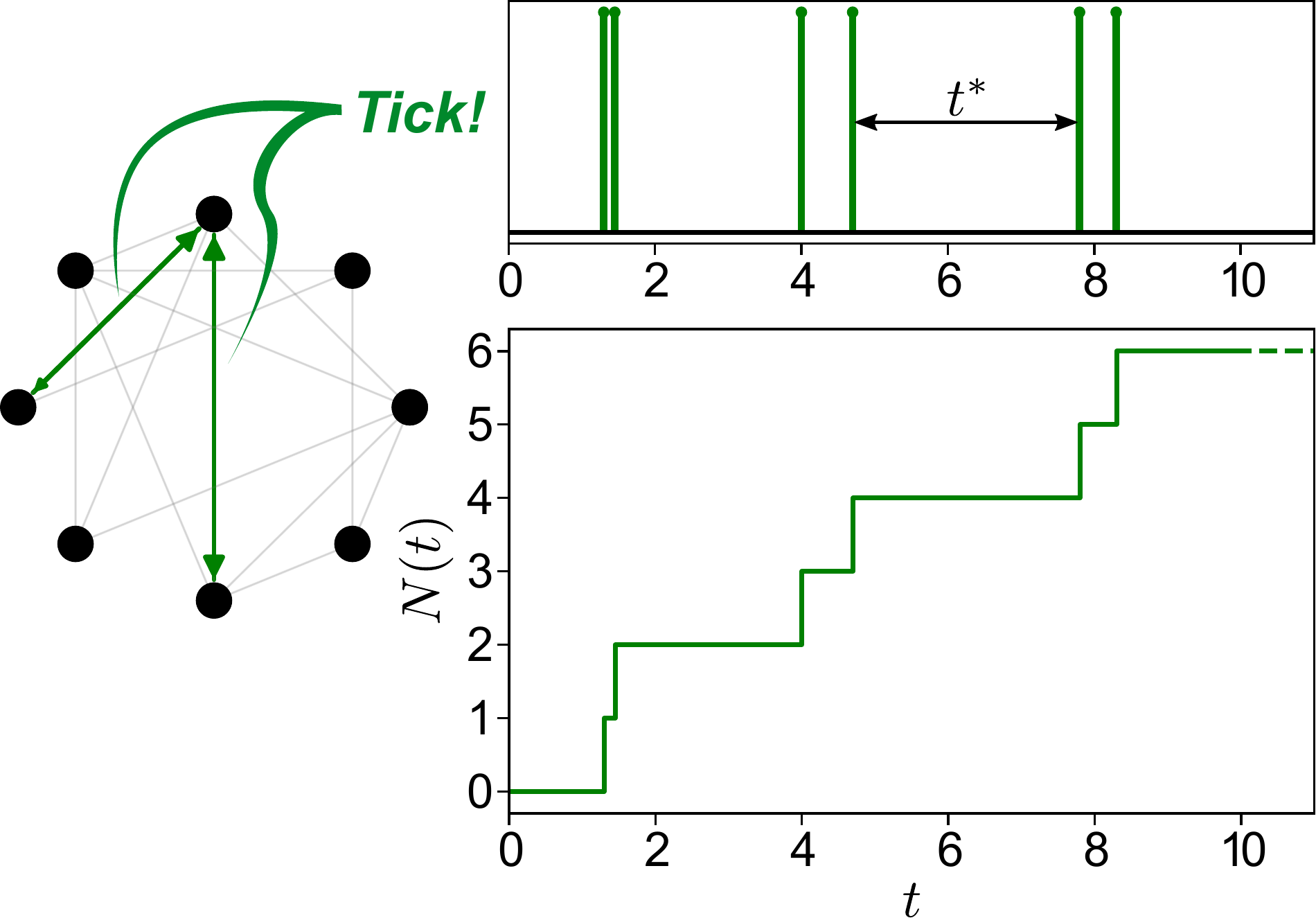}
    \caption{On the left, an exemplary network representing a physical system composed of 8 microscopic states randomly connected to each other. Microscopic states and gray connections are hidden from an external observer. The only information that can be gathered on the system refers to transitions through green (visible) edges. Such transitions (ticks) happen at random times and are represented as vertical spikes in the top-right plot. The total number of spikes is counted over time, labeled $N(t)$, and plotted in the bottom-right plot. Waiting times $t^*$ and total number of ticks $N(t)$ are random variables whose statistics give information about the underlying (mostly inaccessible) process.}
    \label{fig:IntroImage}
\end{figure}

We model the system inside the black box as a continuous time Markov jump process on a network of microscopic states, an example of which appears on the left of Fig.\ \ref{fig:IntroImage}. Thus, we consider the same class of systems for which also the TUR holds~\cite{bara15,ging16}. The transition rates from any state $i$ to another state $j$ are denoted as $k_{ij}$. They are time-independent, such that the system reaches a stationary state with distribution $\pstat_i$. We denote averages as $\mean{\ldots}$. They can be sampled by repeated experiments with initial conditions drawn from $\pstat_i$, or, equivalently, by pieces of a very long stationary trajectory. The average rate of entropy production in this steady state is given by~\cite{seif12}
\begin{equation}
    \sigma=\sum_{ij}\pstat_{i}k_{ij}\ln\frac{\pstat_ik_{ij}}{\pstat_jk_{ji}}.
    \label{eq:sigmatot}
\end{equation}
The sum runs over all pairs of states of the Markov network with $k_{ij}>0$, microreversibility then implies that also $k_{ji}>0$. If, in the long run, entropy production only occurs in a single heat bath at temperature $T$, then $T\sigma$ is the rate of heat dissipation and also the rate at which energy needs to be supplied to keep the system in the non-equilibrium state (setting Boltzmann's constant $k_\mathrm{B}=1$).

The discrete events we refer to as ``ticks'' are generated autonomously by the system and are represented as vertical spikes over time in the top-right plot of Fig.~\ref{fig:IntroImage}. We will generally assume that a tick comes with some change of the microscopic state of the system. Transitions along certain edges always produce a tick, we identify them by setting the constant $b_{ij}=1$ for such transitions from $i$ to $j$. For all other, ``silent'' transitions we set $b_{ij}=0$. The counting observable $N(t)$ increments by one every time a transition with $b_{ij}=1$ occurs, as illustrated in the bottom-right plot of Fig.\ \ref{fig:IntroImage}. Because of the hidden microscopic dynamics of the system, the ticking process appears non-Markovian to an external observer. Given the average flow $\pstat_i k_{ij}$ from state $i$ to $j$ in the steady state, the average rate of ticks follows as
\begin{equation}
\dN=\sum_{ij}\pstat_ik_{ij}b_{ij}.
\end{equation}
This rate of ticks sets the timescale relevant for an external observer. The appropriate dimensionless quantification of the entropic cost is therefore the average entropy produced per tick,
\begin{equation}
    \bar\sigma \coloneqq \sigma/\dN \, .
    \label{eq:sigmabar}
\end{equation}

Counting observables can be classified by their time-symmetry. If for every $b_{ij}=1$ also $b_{ji}=1$, the resulting counting observable is time-symmetric. This means that the same number of ticks is counted in a forward and in a time-reversed trajectory of the microscopic dynamics. We call this type of counting observable ``traffic-like'', since it picks up the non-directed ``traffic'' (the time-symmetric counterpart of the current~\cite{maes08}) for each edge of the Markov network with $b_{ij}=1$. On the other hand, a generic counting observable has no particular symmetry in $b_{ij}$. We call it ``flow-like'', since it picks up the flow of every directed edge with $b_{ij}=1$. This class of observables includes the possibility of some edges contributing symmetrically and some edges asymmetrically (the proof of our bounds for this case is completed in Appendix~\ref{sec:mixed_obs}).

We consider two different quantifications of the precision of the ticking process. First, the Fano factor
\begin{equation}
    F \coloneqq \lim_{t\to\infty}\frac{\mathrm{Var}[N(t)]}{\langle N(t)\rangle}
    \label{eq:fanodef}
\end{equation}
describes the uncertainty in the number of ticks over large time scales. It is dimensionless and typically nonzero and finite, as both the variance $\mathrm{Var}[N(t)]$ and the mean $\langle N(t)\rangle$ increase linearly in the long run in the steady state.
Second, we consider the uncertainty in the waiting time $t^*$ between two successive ticks, captured by the dimensionless coefficient of variation
\begin{equation}
    \varepsilon^2 \coloneqq \frac{\mathrm{Var}[t^*]}{\langle t^*\rangle^2}.
    \label{eq:cv_wt}
\end{equation}
Except in the case where successive ticks are always uncorrelated, both quantifications typically have different numerical values and characterise complementary aspects of the precision. A skilled drummer playing an almost perfect rhythm of alternating half and quarter notes will be characterised by fairly large waiting time fluctuations (counting every beat of the drum as a tick). Yet, the time alternations do not matter in the long run, leading to a small Fano factor. In contrast, a less skilled drummer playing only quarter notes will be characterised by smaller waiting time fluctuations but a larger Fano factor. Both types of precision come at a minimal thermodynamic cost, which we derive in this article.

\section{Main results}
\label{sec:mainres}

\begin{figure}
    \centering
    \includegraphics[width=\linewidth]{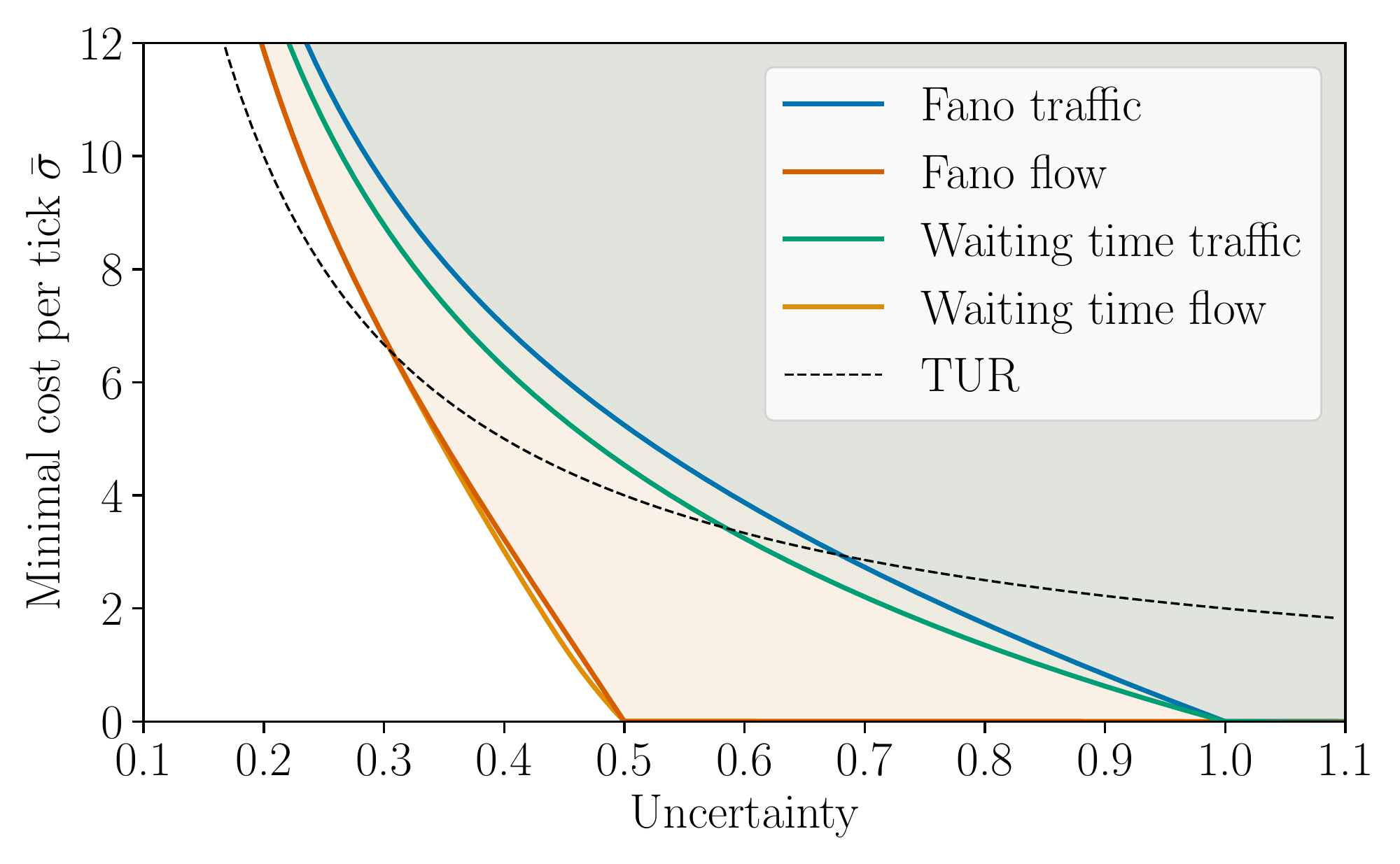}
    \caption{Plot of all four bounds, which constrain possible
      combinations of the entropy production rate per tick and the
      uncertainty of counting observables. The uncertainty on
      the horizontal axis can either be quantified by the Fano factor
      $F$ (blue and orange) or by the coefficient of variation of
      waiting times $\varepsilon^2$ (green and red). The bounds for time-symmetric (``traffic-like'') counting observables
      are shown in blue and green, the ones for generic
      (``flow-like'') observables in orange and red. For comparison, the TUR, which holds for time-antisymmetric observables, is shown as a dashed line.} 
    \label{fig:allbounds}
\end{figure}

We first summarise our four different bounds on the uncertainty, one for each class of counting observable and one for each type of precision. The only parameter in the bounds is the entropy production per tick~$\bar\sigma$. In the cost-uncertainty plane, as shown in Fig.~\ref{fig:allbounds}, the geometric shape of the bounds defines the accessible regions (as so-called Pareto fronts). Hence, they can also be understood as bounds on the entropic cost for an observed or desired precision.

\begin{table*}[t]
\normalsize

\rowcolors{1}{gray!15}{white} 

\begin{tabular}{l !{\vrule width -2.5pt} l !{\vrule width -2.5pt} l !{\vrule width -2.5pt}}
\hline\hline
 & \centering \textbf{Fano factor} & \textbf{Waiting time uncertainty} \\
\hline
\textbf{Traffic} & \Gape[10pt][10pt]{\makecell*[{{p{7cm}}}]{$\displaystyle F\geq \min_x\left[1-x^2+\frac{x^4}{\bar\sigma/2-x\artanh(x)+x^2}\right]$} \eqref{eq:localbound_fano}} & \Gape[10pt][10pt]{\makecell*[{{p{9cm}}}]{$\displaystyle \varepsilon^2\geq\min_x\frac{\bar\sigma(1-x^2)+4x^2-2(1-x^2)x\artanh(x)}{\bar\sigma(1+x^2)+4x^2-2(1+x^2)x\artanh(x)}$ \eqref{eq:localbound_wtd}} } \\
\textbf{Flow} & \Gape[10pt][10pt]{\makecell*[{{p{6cm}}}]{$\displaystyle F\geq\min_x\left[1-\frac{1}{2-x+\frac{2x^2}{\bar\sigma+x\ln(1-x)}}\right]$} \eqref{eq:localbound_flow}} & 
    \Gape[10pt][10pt]{\makecell*[{{p{9cm}}}]{$\varepsilon^2\geq\begin{cases}\displaystyle\min_x\left[1-\frac{1}{2-x+\frac{2x^2}{\bar\sigma+x\ln(1-x)}}\right]\text{ \eqref{eq:localbound_flow}},&\text{if }\bar{\sigma} \geq \bar{\sigma}_c\\[3ex]\displaystyle
    \frac{C'}{(1-C')}\text{ with $C'$ in Eq.~\eqref{eq:localbound_flow_wtd}},&\text{otherwise}\end{cases}$}} \\
\hline\hline
\end{tabular}
\caption{Mathematical form of the bounds on the uncertainty, with references to where they are derived in the main text.}
\label{tab:Bounds}
\end{table*}


All four bounds have a similar mathematical form, as collected in Tab.\ \ref{tab:Bounds}. The bound on the Fano factor for traffic-like and flow-like observables are derived respectively as Eqs.~\eqref{eq:localbound_fano} and \eqref{eq:localbound_flow} in Sec.~\ref{sec:typ}. For the uncertainty in the waiting time for traffic-like observables, the bound follows in Sec.~\ref{sec:wtd} as Eq.~\eqref{eq:localbound_wtd}. For flow-like observables, the bound on waiting time fluctuations has the same mathematical form as the bound on the Fano factor~\eqref{eq:localbound_flow} if the entropy production is above a critical value $\bar\sigma_c\simeq 7.2$. Below this critical value, the bound differs slightly, as discussed in Sec.~\ref{sec:phasetrans}.

Typically, the bounds can be approached asymptotically for Markov networks where only a single, ``visible'' edge with $b_{ij}=1$ (and possibly $b_{ji}=1$) contributes to the counting observable. As discussed in detail in Sec.~\ref{sec:optimal}, the optimal networks consist of a single cycle with a large number of states and uniformly biased rates, except in the vicinity of the visible edge. The only exception is encountered for the bound on waiting time fluctuations of flow-like observables, where a behaviour similar to phase separation can yield higher precision at fixed cost $\bar\sigma<\bar\sigma_c$ by combining the flow of several visible edges with different properties. 

The bounds on the uncertainty of counting observables can be compared with the TUR that
holds for a time-antisymmetric observable $X(t)$. Defining the Fano
factor and the scaled entropy production
$\bar\sigma=\sigma/\langle\dot X\rangle$ accordingly, the TUR can be
written as $\bar\sigma\geq2/F$ (dashed line in
Fig.~\ref{fig:allbounds}). The crucial difference is that for counting
observables, the minimal cost reaches zero at a finite value of the
uncertainty. This value is attained in a simple two-state system with
equal forward and backward transition rates. There, the entropy production in
the steady state is $\sigma=0$. The traffic between the two states describes
a Poisson process, for which $F=1$ and $\varepsilon^2=1$. For the
directed flow from one state to the other, a simple
calculation yields $F=1/2$ and $\varepsilon^2=1/2$. Larger
uncertainties can always be realised at zero entropic cost (e.g., in a two-state system with unequal transition rates). However, smaller
uncertainties soon become more costly than in the case of
time-antisymmetric observables, as visible from the intersections in
Fig.~\ref{fig:allbounds}. For very small uncertainties, the minimal cost scales asymptotically with the inverse of the uncertainty, the dominant term being $2/F$ or $2/\varepsilon^2$, just like in the TUR.

Note the importance of our restriction to counting observables with increments
$b_{ij}\in\{0,1\}$. In the case of the two-state system in equilibrium, taking $b_{12}=b_{21}=\lambda\in\mathbb{R}$ leads to $F=\lambda$. This means that
arbitrary precision could be attained at $\sigma=0$. Hence, for generic
time-symmetric observables with arbitrary increments, there can be no model-independent lower bound on the entropy production other than $\sigma\geq 0$. In contrast, the TUR holds for
observables with time-antisymmetric but otherwise arbitrary real increments.

\section{Applications}
\label{sec:applications}
\subsection{Neural spike trains}
\begin{figure}
    \centering
    \includegraphics[width=\linewidth]{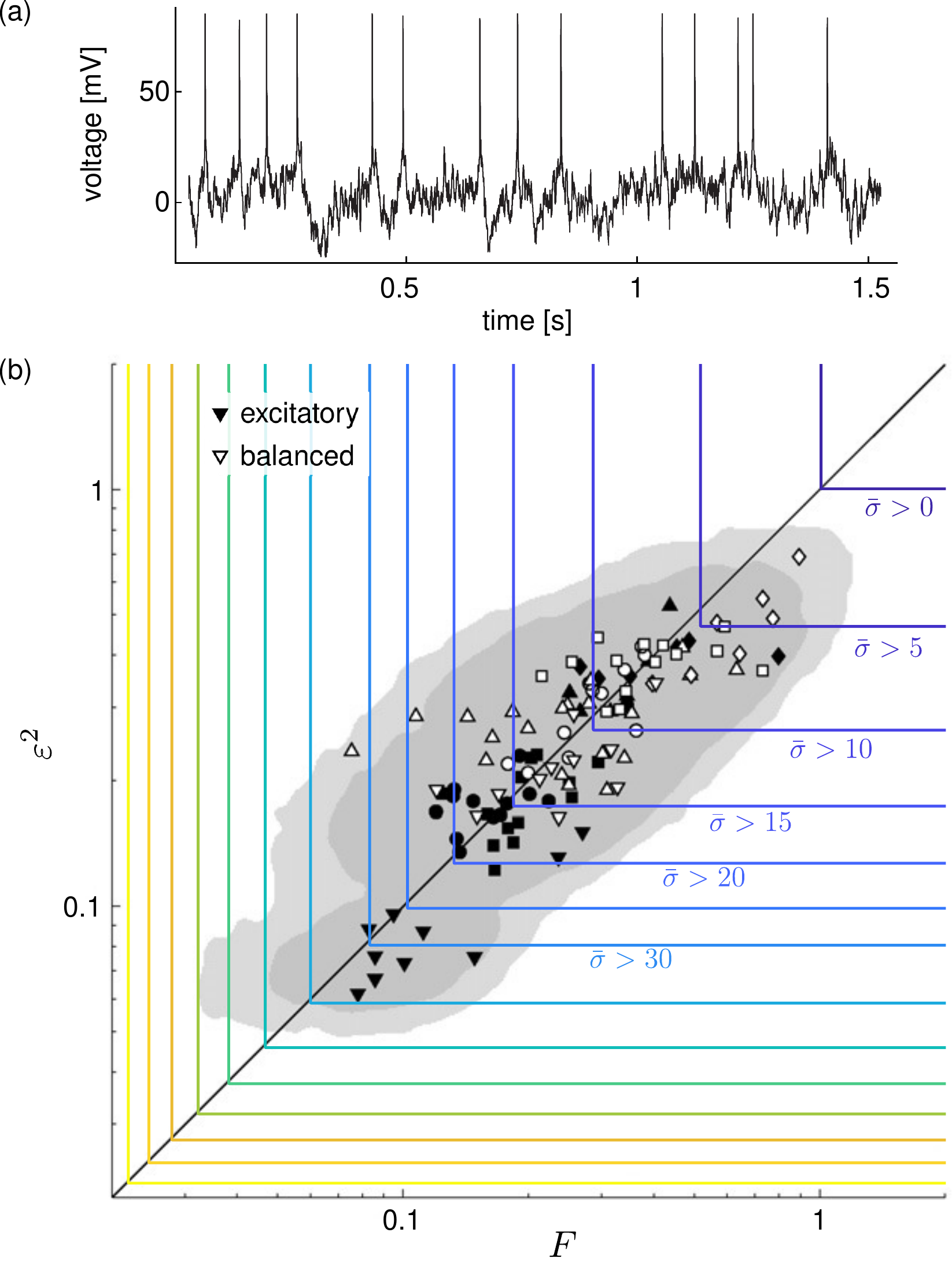} \\
    \caption{(a) Typical time trace for the membrane potential of a neuron, stimulated by Poisson noise~\cite{nawr08}. Short-lived spikes are clearly visible. (b) Characterisation of the variability of the spike train through the Fano factor $F$ and the coefficient of variation of inter-spike intervals $\varepsilon^2$ (data from Ref.~\cite{nawr08}). Solid symbols correspond to purely excitatory stimulation of neurons, open symbols to a mix of excitatory and inhibitory stimulation. Symbol shapes correspond to five different neurons. Using the bounds~\eqref{eq:localbound_fano} and \eqref{eq:localbound_wtd}, every point in the plane can be associated with a minimal entropy production per spike. This minimal entropic cost (measured in $k_\mathrm{B}$ per spike) is indicated through an overlay of isolines. Every point lying left or below one of these lines is associated with at least the amount of entropy production indicated in the label. For details on the uncertainty indicated by the grey shaded regions, see Ref.~\cite{nawr08}.}
    \label{fig:NeuronData}
  \end{figure}
The Fano factor and the coefficient of variation of waiting times are well established as measures of the variability in observed sequences of discrete events. Our bounds add to this statistical characterisation a thermodynamic interpretation through a minimal entropic cost for the underlying microscopic process.

Neurons communicate through temporal patterns in their membrane potential, known as spike trains, whose statistical properties play a major role in neuroscience~\cite{Bialek1991,Bialek1993,Bialek1996,Dayan2001,Stein2005,Faisal2008,nawr08,Rajdl2020,Hasegawa2022}. 
In Fig.~\ref{fig:NeuronData}, we show data for the variability of spike trains recorded in a series of \textit{in vitro} experiments using neurons from the sensorimotor cortex of rats~\cite{nawr08}. The neurons get stimulated through artificially generated excitatory and inhibitory pulses, in a time sequence that is Poissonian distributed. The input signal hence carries no timing information that would go beyond that of an equilibrium bath. Fig.~\ref{fig:NeuronData}a shows the typical measured time evolution of the membrane potential of a single neuron. It exhibits clearly defined spikes, where the potential briefly rises to about 50\,mV. The duration of each of the spikes is very short compared to the time between spikes, such that every rise of the potential above 50\,mV and the subsequent fall can be regarded as a single time-symmetric event.

Fig.~\ref{fig:NeuronData}b shows the Fano factor and the coefficient of variation of inter-spike time intervals obtained from these spike trains. Without any further model assumption about the neurons, we can use our bounds to associate these statistical measures with a minimal entropy production per spike. For balanced excitatory and inhibitory stimulation, the average Fano factor is $0.34$, and the average coefficient of variation is $0.32$. Via the bounds~\eqref{eq:localbound_fano} and~\eqref{eq:localbound_wtd}, these levels of precision correspond to rate of entropy production of at least $8\,k_\mathrm{B}$ per spike. For purely excitatory stimulation, the average values of $0.23$ for the Fano factor and $0.21$ for the coefficient of variation of the inter-spike time interval correspond to the entropy production of at least $12\,k_\mathrm{B}$ per spike. These bounds being larger than zero shows that the process underlying the timing of the spiking neuron must be an active one. Moreover, we see that the processing of the excitatory stimulus comes at a higher minimal energetic cost than the processing of the balanced stimulus (for roughly equal average spike rates~\cite{nawr08}). Note that since we do not assume any renewal property of the spiking process, the measurements of the Fano factor and the coefficient of variation, as well as the corresponding bounds are independent of each other. The contour lines in Fig.~\ref{fig:NeuronData} indicate the larger of the two bounds. For the experimental data at hand, both bounds yield roughly the same entropy production.

At body temperature, these fundamental bounds on entropy production correspond to an energy expenditure of at least $10^{-20}\,\mathrm{J}$ per spike, which is much less than the cost of $10^{-11}\,\mathrm{J}$ estimated for the overall mechanism underlying the generation of a spike~\cite{attw01,poon11}. Hence, one can conclude that most of the energy is spent on the production of the spikes themselves, while the precision in their timing is not a bottleneck for efficient neural information processing.

\subsection{Autonomous clocks}
Our bounds can also be used to interpret the measurements of cost and precision of clocks~\cite{bara16,erke17,pear21,piet22,meie23,gopa23}. Most types of clocks generate a precise sequence of ticks, which are then counted to produce the reading of the measured time. While the counting process is itself noisy and comes at an additional entropic cost~\cite{piet22}, it can be treated as external to the ticking process we focus on here. In Ref.~\cite{erke17}, precision is quantified by the number of ticks $\mathcal{N}$ before the clock is off by one tick on average. For a sufficiently good clock, this will happen after a long time, where the Fano factor~\eqref{eq:fanodef} is then given by $F=1/\mathcal{N}$. Moreover, the high precision of a good clock leads to the special case where the bound is dominated by the leading term of its asymptotic expansion for small $F$ (see Appendix~\ref{sec:parametric}). It then reads $\bar\sigma\geq 2\mathcal{N}$. This bound on the entropy production per tick is the same as the bound formulated in Ref.~\cite{erke17}, which we thus prove rigorously for the whole class of systems that can be modelled as Markov jump processes. This includes discrete quantum systems where decoherence times are much shorter than other relevant time scales. Conversely, this bound may be broken in systems where coherent oscillations (either classical \cite{piet22,gopa23} or quantum~\cite{bran18,guar19,wood22}) are present.

The entropy production $\sigma$ is linked directly to the power consumption, if the clock can be considered autonomous. That means it is only driven by a single external work reservoir (e.g., a battery) or heat reservoir (as in Ref.~\cite{erke17}), and it dissipates only heat into the thermal environment of the observer. Under this condition, the dissipated heat $\sigma T$ equals the input power in the stationary state. Any oscillator (e.g., to probe the hyperfine transition of caesium) or cooling device can be considered an integral part of the clock, whose dissipation is accounted for in $\sigma$.

The measurement of precision and power consumption of an autonomous atomic clock is thus sufficient to determine whether it might be modelled as a Markov jump process as considered here and in Ref.~\cite{erke17}, or whether coherences need to be taken into account. In the context of clocks, the statistical error is quantified by the Allan variance~\cite{alla66}, which in our notation corresponds to $\mathrm{Var}[N(t)]$. The relative uncertainty $\sqrt{\mathrm{Var}[N(t)]}/\mean{N(t)}=\sqrt{F/(\dN t)}$ decays with the averaging time $t$, until statistical errors reach the same level as systematic uncertainties. For state-of-the-art caesium fountain clocks, this uncertainty has been reported as $2.1\times10^{-13}\sqrt{\mathrm{s}/t}$, for an averaging time $t$ of up to 25 days~\cite{wang23}. With the bound $\bar\sigma F\geq 2k_\mathrm{B}$, this would correspond to the heat dissipation of at least $\bar\sigma T\dN=190\,\mathrm{kW}$ in a thermal environment of $T=300\,\mathrm{K}$. If the experimental setup uses less energy than that, then the design of the clock must make use of coherences. The case for such coherences is even clearer for the current record holders in precision, optical lattice clocks, with a reported relative uncertainty of $4.4\times10^{-18}\sqrt{\mathrm{s}/t}$~\cite{both22}. Without coherences, this would correspond to a power of $430\,\mathrm{TW}$, spending a significant fraction of the world's yearly energy usage during the measurement time of just 92 hours. As a final example, chip-scale atomic clocks~\cite{kitc18} with a reported relative uncertainty of $4\times10^{-11}\sqrt{\mathrm{s}/t}$~\cite{knap05} would need to dissipate at least $5\,\mathrm{W}$ into the thermal environment at $295\,\mathrm{K}$. The reported total power consumption of just $195\,\mathrm{mW}$ is again a clear indicator of coherences.

Beyond the context of clocks, a common measure for fluctuations of counting observables in quantum systems is Mandel's parameter $Q$, related to the Fano factor as $Q=F-1$ \cite{mand79,garr10,mani23}. Hence, measurements of this parameter along with measurements of the entropy production can likewise be used to detect coherences.

\section{Illustration}
\label{sec:illustration}

In this Section, we analyse the tightness of the bounds for examples of concrete Markov networks and counting observables.  For this purpose, we need exact expressions for the precision, as opposed to the variational principles from which the bounds are derived.

\subsection{Exact calculations for general networks}
\label{subsec:Fano}

We consider a network of $M$ states connected to each other through a certain set of edges and characterised by a transition-rate matrix $\boldsymbol{L}$ with components $L_{ij} = k_{ij}$ for $i\neq j$ and $L_{ii}=-\sum_{j} k_{ij}$. 
The scaled cumulant generating function (SCGF) associated with a counting observable $N(t)$ is defined as
\begin{equation}
    \Psi(s)\coloneqq\lim_{t\to\infty}\frac{1}{t}\ln\mean{e^{sN(t)}}.
    \label{eq:scgf_def}
\end{equation}
The Fano factor \eqref{eq:fanodef} can then be obtained as 
\begin{equation}
    \label{eq:FanoSCGF}
    F = \frac{1}{\dN}\partial_s^2\Psi(s)\Bigr|_{s=0} \ .
\end{equation}
The SCGF can be shown to be equal to the dominant eigenvalue (i.e., the eigenvalue with the largest absolute value, which is real) of the so-called tilted matrix $\tilde{\boldsymbol{L}}(s)$ with components~\cite{Ellis1985,touc09,jack10}
\begin{equation}
    \tilde L_{ij}(s)=L_{ij}e^{b_{ij}s}.
    \label{eq:TiltedMatrix}
\end{equation}
We identify the corresponding left eigenvector as $\boldsymbol{l}(s)$ and right eigenvector as $\boldsymbol{r}(s)$,
\begin{equation}
    \boldsymbol{l}(s)\tilde{\boldsymbol{L}}(s)=\Psi(s)\boldsymbol{l}(s)\ \mathrm{and}\ \tilde{\boldsymbol{L}}(s)\boldsymbol{r}(s)=\Psi(s)\boldsymbol{r}(s),
    \label{eq:ev}
\end{equation}
and choose the normalisation 
\begin{equation}
    \label{eq:ev_norm}
    \sum_il_i(s)=1 \textup{ and }\sum_il_i(s)r_i(s)=1. 
\end{equation}
Note that, by construction, $l_i(0)=\pstat_i$ and $r_i(0)=1$ for all $i$.
Cumulants of the counting observable such as in \eqref{eq:FanoSCGF} can be obtained algebraically from an expansion of eigenvalues and -vectors in $s$~\cite{baie09}. Alternative methods for calculating such cumulants are based on the characteristic polynomial of the tilted matrix~\cite{koza99} or graph-theoretic considerations~\cite{padm23}.

The matrix $\boldsymbol{S}\coloneqq\tilde{\boldsymbol{L}}(-\infty)$ describes the dynamics where each transition with $b_{ij}=1$ leads to an absorbing state. With that, the waiting time distribution can be calculated as~\cite{Rubino1989,seki21,skin21,haru22}
\begin{equation}
    p(t^*)= \frac{1}{\dN} {\boldsymbol{p}\stat\boldsymbol{S}e^{t\boldsymbol{S}}\boldsymbol{S}\boldsymbol{1}} \, ,
\end{equation}
using the matrix exponential and the vector $\boldsymbol{1}$ containing $M$ ones. The uncertainty in the waiting time~\eqref{eq:cv_wt} follows as
\begin{equation}
    \varepsilon^2=-2\dN\boldsymbol{p}\stat\boldsymbol{S}^{-1}\boldsymbol{1}-1.
    \label{eq:cv_exact}
\end{equation}

\subsection{Numerical illustration}

\begin{figure}
    \centering
    \includegraphics[width=\linewidth]{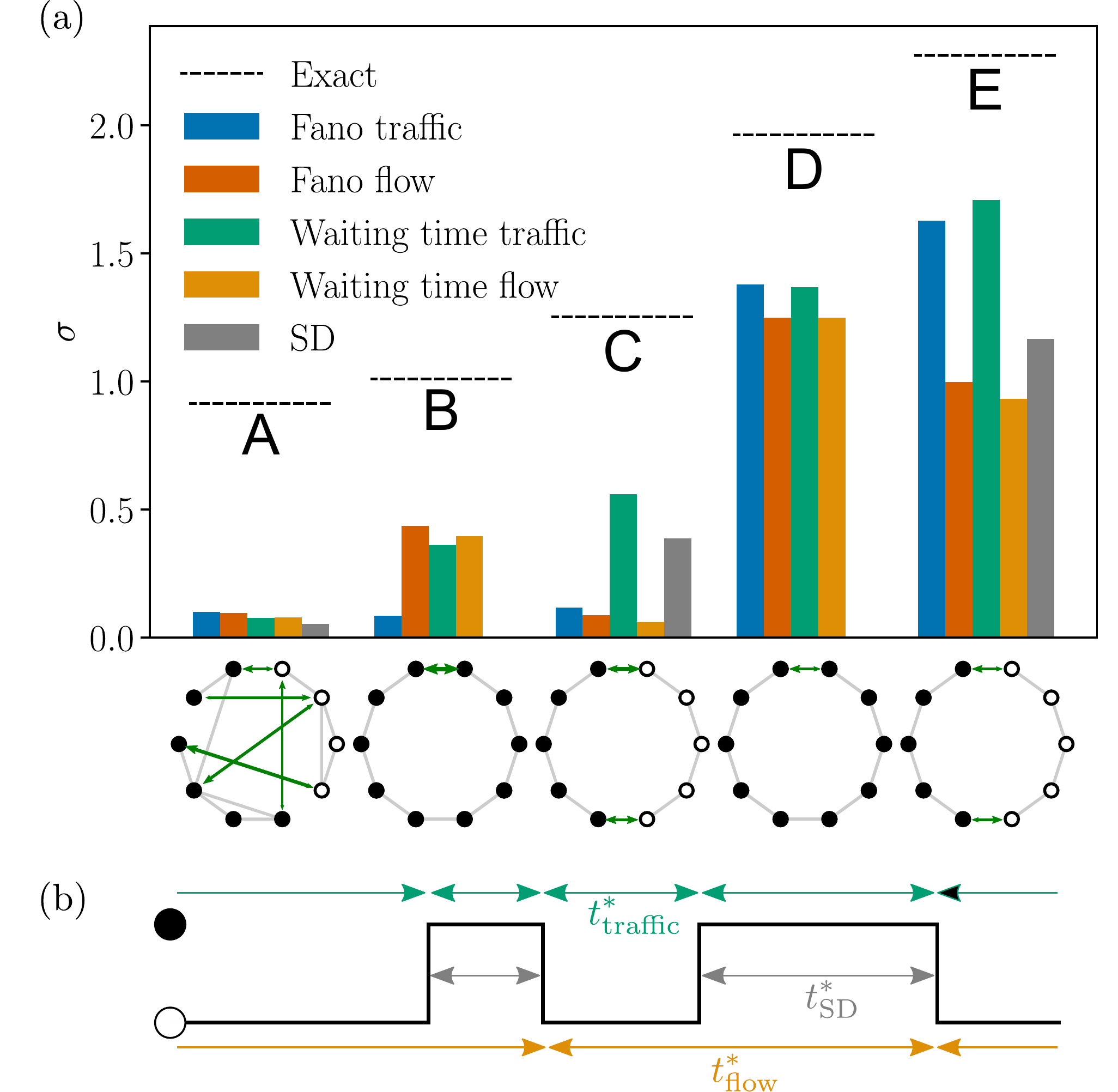}
    \caption{(a) Inference of the entropy production rate $\sigma$ from the precision of a counting observable using the four bounds of Fig.~\ref{fig:allbounds}. The schematic network structure is indicated below each set of columns. Edges contributing via their traffic or flow to the counting observable are coloured in green and the thickness of arrows indicates the magnitude of transition rates. Dashed lines represent the exact entropy production and coloured columns the lower bounds inferred from the precision. Where the network states can be grouped in two classes (black and white) connected only by visible edges, our bounds can be compared to the one by Skinner and Dunkel (SD, grey)~\cite{skin21}. (b) Identification of the waiting times entering our bounds ($t^*_\mathrm{traffic}$, $t^*_\mathrm{flow}$) and the bound of Ref.~\cite{skin21} ($t^*_\mathrm{SD}$) for an example of a time series recorded by an observer who can only distinguish the two classes (black and white) of states.}
    \label{fig:SingleNets}
\end{figure}

In order to evaluate the tightness of our bounds, we now calculate numerically the rate of entropy production~\eqref{eq:sigmatot} and the measures for precision~\eqref{eq:FanoSCGF} and \eqref{eq:cv_exact} for examples of Markov networks. 

In Fig.~\ref{fig:SingleNets}(a), we show the results for five different examples. First, network A has many connections with random transition rates. We divide the states in two subsets, black and white. A traffic-like observable is defined by counting any transition between black and white as a tick, for a flow-like observable we count only directed transitions from black to white. In the vast majority of such random networks, all measures of precision are greater than $1$, making our bounds not more informative than the 2nd law $\sigma\geq 0$. Out of a random generation of networks, less than 1\,\% of them yield any non-zero bound, the results for one of them are shown as the columns labeled A in Fig.~\ref{fig:SingleNets}(a). Still, we infer only a small fraction of the actual entropy production. Nevertheless, such bounds play out their strengths for technologically or biologically relevant systems that have been designed or have evolved to produce periodic oscillations.

The networks B and C of Fig.~\ref{fig:SingleNets}(a) are unicyclic, with random, unbiased transition rates. One (B) or two (C) transitions contribute to the counting observable. Here, the lower estimate of the entropy production is typically non-zero, but still less than half of the actual value. The best results are typically obtained for random transition rates that are biased in a single direction (D and E). This is also the type of setting where the TUR is typically strong (being saturated by a unicyclic network with equal forward and equal backward rates with a slight bias).

It is worth noting that in each scenario, different bounds give the best estimate on $\sigma$. For the experimental inference of the entropy production, it is therefore advisable to consider different types of observables (if available) and both types of precision and pick the largest lower bound on $\sigma$. 

In networks A, C, and E, we count the transitions between two classes of states. This corresponds to an experimental setting where an observer has access to a single binary observable, with a trajectory as shown in Fig.~\ref{fig:SingleNets}(b). Such an observation could come, for example, from  a molecule switching between a fluorescent and a non-fluorescent state. This setting is the same as the one considered by Skinner and Dunkel in Ref.~\cite{skin21}. Their bound on $\sigma$, obtained numerically through optimisation of finite networks, uses the fluctuations of the waiting time between entering and leaving one of the classes. This is conceptually similar but different from the waiting times considered in our bounds, as illustrated in Fig.~\ref{fig:SingleNets}(b). Even though our bounds are applicable to a wider range of settings (including B and D), they can give stronger estimates of $\sigma$ than the one of Ref.~\cite{skin21}.

\begin{figure}
    \centering
    \includegraphics[width=\linewidth]{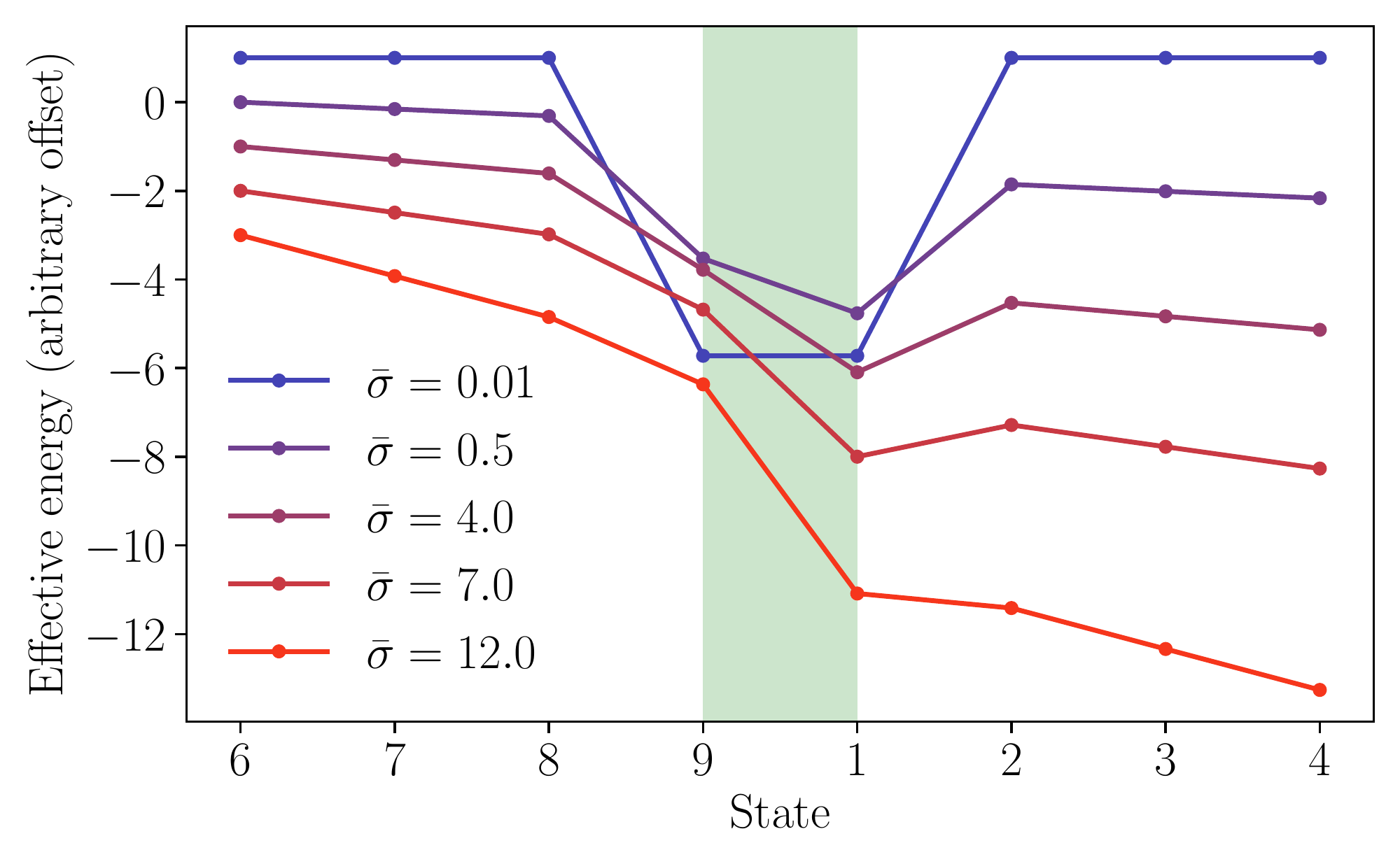}
    \caption{Optimal energy landscape around the visible edge of a unicyclic network with $M=9$ states. This landscape minimises the fluctuations of the traffic between states $9$ and $1$ at given scaled entropy production $\bar\sigma$. The ``effective energy'' represents both the free energy of a state and the non-conservative force driving the system, indicated through the overall slope. Its difference is constant along all edges, except for the observed one and its two neighbours.}
    \label{fig:OptimalLandscape}
\end{figure}

Inspired by the observation that the bounds on entropy production are typically tightest for unicyclic networks, we restrict our search for optimal networks within this class. Moreover, we focus on a single visible edge, as in the derivation of the previous Sections. For notational convenience, for a unicyclic network of $M$ states we now label the visible edge as $(M,1)$.
A numerical optimisation of the transition rates to minimise traffic fluctuations for a given scaled entropy production reveals an interesting pattern: All forward rates are equal and all backward rates are equal, except for the transitions of the visible edge $(M,1)$ and the other two transitions leading out of states $M$ and $1$. The local detailed balance ratio $\ln(k_{ij}/k_{ji})$ between two neighbouring states can be interpreted as a change in effective energy, involving the free energy difference between the states and the work done by a non-conservative driving force~\cite{seif12}. This adds up to an effective energy landscape as shown in Fig.~\ref{fig:OptimalLandscape} (Note the mismatch in this effective energy extrapolated to state 5 on the left and right side of the plot, which is due to the work accumulated over one cycle.) The visible edge is embedded in a potential well that gets increasingly distorted with increasing $\bar\sigma$. Since the change in energy for the visible edge is steeper then elsewhere, frequent re-crossings of the visible edge that would spoil the precision are suppressed. For the bulk of edges the slope is more shallow, in order to reduce the thermodynamic cost of driving. In Sec.~\ref{sec:optimal}, we will calculate the cost and precision exactly for this type of network and show that our bounds can be saturated.

\section{Large deviation bounds for a single visible edge}
\label{sec:ldf}

We start with the derivation of bounds on the large deviation function, capturing the distribution $p(N,t)$  of a counting observable in the long-time limit~\cite{touc09}. As the mean value $t\dN$ becomes typical for $N(t)$, the probability of fluctuations away from the mean by a factor of $\nu$ (rounding to a nearby integer $\lfloor \nu t\dN\rfloor$) decays exponentially, as described by the large deviation function
\begin{equation}
    I(\nu) \coloneqq -\lim_{t\to\infty}\frac{1}{t}\ln\left[p(N=\lfloor \nu t\dN\rfloor,t)\right].
    \label{eq:ldfdef}
\end{equation}
This function is related to the SCGF~\eqref{eq:scgf_def} through a Legendre transform~\cite{touc09}.
Due to the non-negativity of $N$, also $I(\nu)$ is only defined for $\nu\geq0$. We define $\bar I(\nu) \coloneqq I(\nu)/\dN$ as a dimensionless rate function.  It will turn out that, similar to the bounds on current fluctuations~\cite{piet16}, $\bar I(\nu)$ can be bounded from above by a function depending only on the entropy production $\bar\sigma$.

We employ the formalism of level 2.5 large deviation theory for Markov jump processes \cite{De2001,maes08,Bertini2012,Barato2015}, on which also the original proof of the TUR was based~\cite{ging16}. It considers the empirical flow $n_{ij}$, which is the fluctuating number of transitions from $i$ to $j$ divided by the observation time $t$. Its mean value in the stationary state is $n\stat_{ij}=\pstat_i k_{ij}$. Moreover, the empirical density $p_i$ is the fraction of the observation time spent in state $i$, with mean $\pstat_i$.
The large deviation function for the joint distribution of all $n_{ij}$ and all $p_i$, defined analogously to Eq.~\eqref{eq:ldfdef}, is given by~\cite{maes08}
\begin{equation}
    I(\{p_i\},\{n_{ij}\})=\sum_{ij}\left[n_{ij}\ln\frac{n_{ij}}{p_ik_{ij}}-n_{ij}+p_ik_{ij}\right],
    \label{eq:lvl25general}
\end{equation}
with the sum running over all edges $(ij)$ with $k_{ij}\neq 0$ and the constraint $\sum_j (n_{ij}-n_{ji})=0$ for all $i$ (otherwise $I$ is formally assigned $\infty$).
    
A general counting observable can be expressed as $N(t)=t\sum_{ij}n_{ij}b_{ij}$. By the contraction principle, $I(\nu)$ can be obtained from $I(\{p_i\},\{n_{ij}\})$ by minimising over all $p_i$ and $n_{ij}$ with fixed $\sum_{ij}n_{ij}b_{ij}=\nu\dN$. A bound on $I(\nu)$ can be readily obtained by using a suboptimal ansatz for $p_i$ and $n_{ij}$ consistent with the constraints.

The flow $n_{ij}=(j_{ij}+\traf_{ij})/2$ of a directed edge of a Markov network can be decomposed into the time-antisymmetric current 
\begin{equation}
 j_{ij} \coloneqq n_{ij}-n_{ji}   
\end{equation}
and the time-symmetric traffic 
\begin{equation}
\traf_{ij} \coloneqq n_{ij}+n_{ji}    
\end{equation}
(and likewise for the stationary averages $n\stat_{ij}$, $j\stat_{ij}$, and $\traf\stat_{ij}$). The bound on the large deviation function for time-antisymmetric observables, which led to the original proof of the TUR, was derived in the following three steps~\cite{ging16}: First, a full contraction over the traffic of every edge with the optimal value
\begin{equation}
    \traf_{ij}^*=\sqrt{4p_ik_{ij}p_jk_{ji}+j_{ij}^2},
    \label{eq:trafopt}
\end{equation}
then taking the ansatz with $p_i=\pstat_i$ and all currents scaled to $j_{ij}=\xi j\stat_{ij}$ by a common factor $\xi\in \mathbb{R}$, and finally bounding the result from above by a quadratic function. A similar ansatz was used in Ref.~\cite{harv20}, but with the empirical distribution $p_i$ kept variable to derive a thermodynamic bound on the fluctuations of the time spent in a set of states---a time-symmetric observable complementary to the traffic we consider here.

\subsection{Traffic fluctuations}
\label{sec:ldf_traffic}

We focus at first on the case of counting observables where ticks are produced in a single edge of the Markov network. Without loss of generality, we label the two states linked by this visible edge as $1$ and $2$. In the time-symmetric case we then have $b_{12}=b_{21}=1$ as the only non-zero $b_{ij}$, such that the counting observable becomes the integrated traffic $N(t)=\traf_{12}t$.

As the crucial next step, we choose an ansatz similar to Ref.~\cite{ging16}. That is, we choose $p_i=\pstat_i$ for the empirical density and $j_{ij}=\xi j\stat_{ij}$  for the current of all edges. For the traffic, we single out the edge $(1,2)$ that is constrained by the counting observable, setting $\traf_{12}=\traf_{21}=\nu\traf\stat_{12}$. For all other edges, we use the optimal value $\traf_{ij}=\traf_{ij}^*$ of Eq.~\eqref{eq:trafopt}. Plugging this choice for the empirical density and the resulting flow into Eq.~\eqref{eq:lvl25general} yields an upper bound on the large deviation function depending on the two parameters $\nu$ and $\xi$. 

Making use of the symmetry properties of $\traf_{ij}$ and $j_{ij}$, Eq.~\eqref{eq:lvl25general} can be written as a sum over unique edges, labeled $(ij)$ with $i<j$. For all edges other than $(1,2)$, our ansatz yields the same terms as in Ref.~\cite{ging16}, which, as shown  there, can be bounded further by $\sigma_{ij}(\xi-1)^2/4$. Here, $\sigma_{ij} \coloneqq j\stat_{ij}\ln(n\stat_{ij}/n\stat_{ji})$ is the entropy production associated with the edge $(ij)$. It is non-negative and contributes to the overall entropy production $\sigma=\sum_{i<j}\sigma_{ij}$, equal to Eq.~\eqref{eq:sigmatot}. With that, we obtain the bound
\begin{align}
    I(\nu)\leq \min_\xi\Bigg[&\sum_{i<j\backslash (1,2)}\frac{\sigma_{ij}}{4}(\xi-1)^2-\traf_{12}+\traf\stat_{12}\nonumber\\
    &+\sum_\pm\frac{\xi j\stat_{12}\pm\nu\traf\stat_{12}}{2}\ln\frac{\xi j\stat_{12}\pm\nu\traf\stat_{12}}{j\stat_{12}\pm\traf\stat_{12}}\Bigg],   
    \label{eq:ldf_traffic0}
\end{align}
where the first sum runs over all edges other than $(1,2)$. The remaining terms come from the edge $(1,2)$, where the two possible directions indicated by $\pm$ are summed over. The minimisation over $\xi$ is a contraction over the scaled currents, which can be varied independently of the counting observable.

The stationary traffic $\traf\stat_{12}$ of Eq.~\eqref{eq:ldf_traffic0} is just the average tick rate $\dN$ and thus accessible to the external observer. In contrast, the stationary current $j\stat_{12}$ would require insights into the microscopic details, or at least a distinction between ticks associated with forward and backward transitions. For a dimensionless characterisation of this current, we write $x \coloneqq j\stat_{12}/\traf\stat_{12}$. Note that, by definition, $|x|<1$. The entropy production of the edge $(1,2)$ can then be written as $\sigma_{12}/\dN=x\ln[(1+x)/(1-x)]=2x\artanh(x)$, and the entropy production summed over all other edges follows as $\sigma-\sigma_{12}$. We scale all terms in Eq.~\eqref{eq:ldf_traffic0} by $\dN$ to arrive at
\begin{equation}
  \bar I(\nu)\leq\max_{x}\min_\xi B_\mathrm{t}(\bar\sigma,x,\xi,\nu)
  \label{eq:Itraf}
\end{equation}
with
\begin{align}
  B_\mathrm{t}(\bar\sigma,x,\xi,\nu) \coloneqq &\frac{1}{4}\left(\bar\sigma-2x\artanh x\right)(\xi-1)^2+1-
    \nu\nonumber\\
    &+\frac{\nu+\xi x}{2}\ln\frac{\nu+\xi x}{1+x}+\frac{\nu-\xi x}{2}\ln\frac{\nu-\xi x}{1-x},
    \label{eq:Btraf}
\end{align}
where the subscript `t' stands for traffic.
In the case of unknown $x$ we need to take the weakest possible upper bound on $\bar I(\nu)$, hence the maximisation over $x$ in Eq.~\eqref{eq:Itraf}. To ensure that the entropy production of the visible edge is less than the total entropy production, the range of $x$ is constrained by $2x\artanh x\leq\bar\sigma$. 

\begin{figure}
    \centering
    \includegraphics[width=\linewidth]{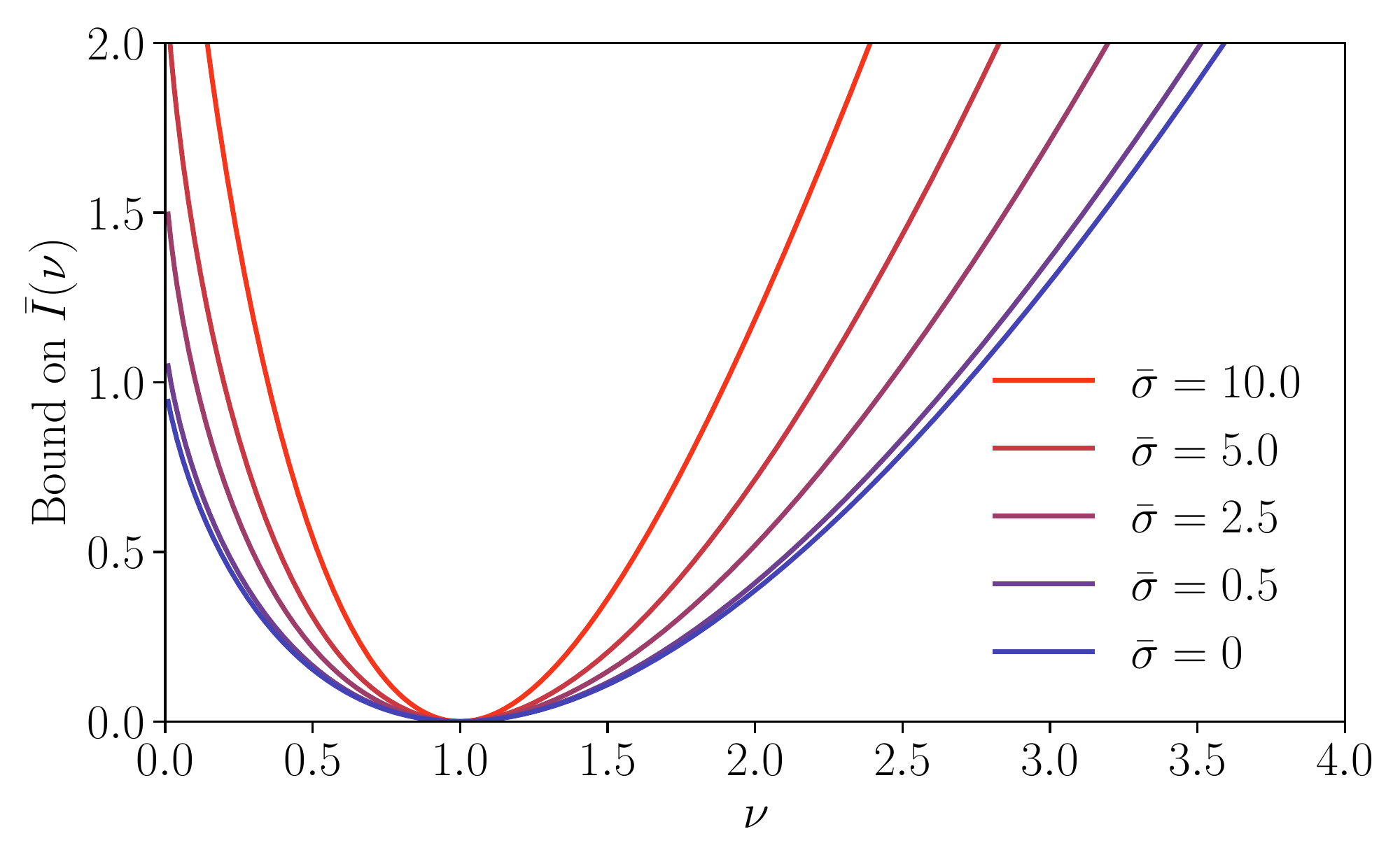}
    \caption{The upper bound \eqref{eq:Itraf} on the large deviation function $\bar I(\nu)$, evaluated for selected values of $\bar\sigma$.}
    \label{fig:LargeDeviationBounds}
\end{figure}

The optimisation over $\xi$ and $x$ can be performed numerically. As shown in Fig.~\ref{fig:LargeDeviationBounds}, the upper bound on $\bar I(\nu)$ gets narrower with increasing entropy production $\bar\sigma$, thus allowing for higher precision.

In the special case of an equilibrium system, where $\bar\sigma=0$, the only
allowed value for $x$ is $x=0$. The bound~\eqref{eq:Itraf} then reduces to
\begin{equation}
  \bar I(\nu)\leq 1-\nu+\nu\ln\nu.
  \label{eq:poisson}
\end{equation}
The right hand side is the large deviation function of a Poisson process. This bound
can be saturated by a simple two-state system with equal transition rates.

\subsection{Flow fluctuations}
We now turn to counting observables consisting of the flow of a single edge $(1,2)$ with $b_{12}=1$ and $b_{21}=0$, i.e., $N(t)=n_{12}t$. The ansatz for the empirical density and flow is essentially the same as in the case of the traffic observable above. The only difference is that for the flow of the visible edge, we choose $n_{12}=\nu n\stat_{12}$ and $n_{21}=n_{12}-j_{12}=\nu n\stat_{12}-\xi j\stat_{12}$. For a dimensionless quantification of the current we now define $y \coloneqq j\stat_{12}/n\stat_{12}$, which is $-\infty<y<1$. The entropy production of the observed link becomes $\sigma_{12}/\dN=-y\ln(1-y)$. 

Following similar steps as for the traffic above, we obtain the bound
\begin{equation}
  \bar I(\nu)\leq\max_y\min_\xi B_\mathrm{f}(\bar\sigma,y,\xi,\nu),
  \label{eq:Iflow}
\end{equation}
where now
\begin{align}
  B_\mathrm{f}(\bar\sigma,y,\xi,\nu) \coloneqq &\frac{1}{4}[\bar\sigma+y\ln(1-y)](\xi-1)^2+2-2\nu\nonumber\\
  &+\nu\ln\nu-y+\xi
  y+(\nu-\xi y)\ln\frac{\nu-\xi y}{1-y} \, ,
  \label{eq:Bflow}
\end{align}
with the subscript `f' standing for flow. The maximisation runs over all $y$ with $-y\ln(1-y)<\bar\sigma$.

In equilibrium, where $\bar\sigma=0$ requires $y=0$, the bound reads $\bar
I(\nu)\leq2(1-\nu+\nu\ln\nu)$, which is twice the large deviation function of
a Poisson process. Again, this bound is saturated for a simple two-state network with equal forward and backward transition rate, with the difference that only
forward transitions are detected. The narrower distribution compared to traffic fluctuations relates to an improvement in precision. It can be attributed to
the additional unobserved resetting step that is necessary between two
successive ticks, making the waiting time distribution sharper than the exponential distribution of a Poisson process.

\section{Typical fluctuations}
\label{sec:typ}

While the large deviation function $\bar I(\nu)$ describes the whole range of fluctuations, including extremely unlikely ones, only typical fluctuations can easily be measured experimentally. Typical fluctuations can be assessed by considering the quadratic expansion of $\bar I(\nu)$ around its minimum at the most likely value $\nu=1$, which yields the Gaussian approximation of the distribution $p(N,t)$. Through the mean and variance of this distribution, the Fano factor~\eqref{eq:fanodef} follows as
\begin{equation}
    F= \left( \partial_\nu^2\left.\bar I(\nu)\right|_{\nu=1} \right)^{-1}.
    \label{eq:Ftyp}
\end{equation}
As a more direct route towards bounds on typical fluctuations, one could limit the discussion to infinitesimal perturbations to the stationary state in the first place~\cite{dech20}.

The discussion that follows applies to both the traffic and to the flow of a single visible edge, taking either $B_\mathrm{t}(\bar\sigma,x,\xi,\nu)$ of Eq.~\eqref{eq:Itraf} or $B_\mathrm{f}(\bar\sigma,y,\xi,\nu)$ of Eq.~\eqref{eq:Iflow} for $B$.
The value of $\xi$ that minimises $B$ for $\nu=1$ is $\xi=1$, which corresponds to the most likely, stationary current $j_{ij}=j\stat_{ij}$. Here, the bound evaluates to $B=0$ and coincides with the large deviation function $\bar I(0)=0$. The curvature of $\bar I$ at $\nu=0$ entering Eq.~\eqref{eq:Ftyp} must therefore be less than that of its upper bound. 

Since we can expect the typical fluctuations of $\nu$ and $\xi$ to be close to $1$, it is useful to expand $B$ to quadratic order as
\begin{equation}
    B\approx\frac{1}{2}\alpha(\xi-1)^2+\beta(\xi-1)(\nu-1)+\frac{1}{2}\gamma(\nu-1)^2.
    \label{eq:quadform}
\end{equation}
The coefficients are
\begin{equation}
\begin{split}
  \alpha &= \frac{\bar\sigma}{2}-x\artanh(x)+\frac{x^2}{1-x^2}, \\
  \beta &= \frac{-x^2}{1-x^2}, \\
  \gamma &= \frac{1}{1-x^2} 
  \label{eq:coeff_traffic}
\end{split}
\end{equation}
for the traffic and
\begin{equation}
\begin{split}
  \alpha &= \frac{1}{2}[\bar\sigma+y\ln(1-y)]+\frac{y^2}{1-y}, \\ \beta &= \frac{-y}{1-y}, \\
  \gamma &= \frac{2-y}{1-y} 
  \label{eq:coeff_flow}
\end{split}
\end{equation}
for the flow. The minimisation over $\xi$ then yields
\begin{equation}
    \min_\xi B=\frac{1}{2}(\gamma-\beta^2/\alpha)(\nu-1)^2+\mathcal{O}[(\nu-1)^3].
\end{equation}
and as bound on the Fano factor~\eqref{eq:Ftyp}
\begin{equation}
    F\geq\min_{x\textup{ or }y}\frac{\alpha}{\alpha\gamma-\beta^2},
    \label{eq:localbound_general}
\end{equation}
minimising over $x$ for the traffic or $y$ for the flow.

Plugging in the coefficients~\eqref{eq:coeff_traffic}, we obtain the bound on the Fano factor of the traffic
\begin{equation}
  F\geq \min_x\left[1-x^2+\frac{x^4}{\bar\sigma/2-x\artanh(x)+x^2}\right].
  \label{eq:localbound_fano}
\end{equation}
Analogously, the coefficients~\eqref{eq:coeff_flow} yield the bound on the Fano factor of the flow
\begin{equation}
  F\geq\min_y\left[1-\frac{1}{2-y+\frac{2y^2}{\bar\sigma+y\ln(1-y)}}\right].
  \label{eq:localbound_flow}
\end{equation}
Both bounds are plotted in Fig.~\ref{fig:allbounds}.

The final minimisation over $x$ or $y$ cannot be performed analytically. Nonetheless, a numerical optimisation can be avoided by plotting the bounds parametrically using the analytic functions $\bar\sigma(x)$ or $\bar\sigma(y)$ of the values of $\bar\sigma$ for which a given value of $x$ or $y$ is optimal, see Appendix~\ref{sec:parametric}. There, we also provide asymptotically exact expressions for small and large $\bar\sigma$.

\section{Waiting time fluctuations}
\label{sec:wtd}

We now turn to the quantification of the precision through fluctuations of waiting times between ticks. Recent research has shown that many results of stochastic thermodynamics can be generalised to evaluation not at a fixed time but at a time determined by the system's dynamics. This applies in particular to first passage times of either the system reaching a certain state~\cite{rold19,pal21} or an integrated observable reaching a certain value~\cite{garr17,ging17,neri17,neri20,hiur21,Neri2022}.

Recall the coefficient of variation $\varepsilon^2=\operatorname{Var}(t^*)/\langle t^*\rangle^2$ [Eq.~\eqref{eq:cv_wt}] of the waiting time $t^*$ between any two successive ticks.
The average is taken over all waiting times between ticks recorded in
a long trajectory.  For the mean waiting time we have the simple
relation $\mean{t^*}=1/\dN$ (see below). However, for the second
cumulant, the relation $F=\varepsilon^2$ holds only in a renewal process \cite{Grimmet2001,Feller2008}, where successive waiting times are uncorrelated. This is typically not the case in a nontrivial Markov network.

To derive bounds on waiting time distributions we employ path reweighting techniques similar to level 2.5 large deviation theory. We denote by $i(t)$ the stochastic trajectory that follows a
tick at time $t=0$ up to the time $t^*$ of the following
tick [the final transition producing this tick is still recorded in $i(t)$]. The time $t^*$ is itself a functional $t^*[i(t)]$ of the trajectory. For the
steady state statistics of these trajectories, we can consider an ensemble of
trajectories $i(t)$ obtained from a very long trajectory of duration $t_\mathrm{obs}$ involving many ticks. We then cut the long trajectory
into pieces at every tick and shift the start time to zero for every snippet. 
By the law of large numbers, the number of snippets is $N_\mathrm{obs}=\dN t_\mathrm{obs}$. On the other hand, since the total time is the sum of all waiting times, we get $t_\mathrm{obs}=N_\mathrm{obs}\mean{t^*}$, which yields $\mean{t^*}=1/\dN$.

The probability of observing a particular $i(t)$ in the ensemble of snippets is
given by the path weight
\begin{equation}
  p[i(t)]=p_{i(0)}(0)\exp\left(-\sum_iw_it_i[i(t)]\right)\prod_{i,j}k_{ij}^{n_{ij}[i(t)]}.
  \label{eq:pathweight}
\end{equation}
Here, $t_i[i(t)]$ is the time spent in state $i$ and $n_{ij}[i(t)]$ is the
number of directed transitions from $i$ to $j$ (including the transition associated with the tick at the end of the trajectory but not
the one at its beginning). The exit rate out of state $i$ is
$w_i \coloneqq \sum_j k_{ij}$.  The initial distribution $p_{i(0)}(0)$ is the probability to find the
system in state $i(0)$ immediately after a tick.

The cumulant generating function (CGF) of the waiting time is defined as
\begin{equation}
  \ln\mean{e^{zt^*}}=\ln\sum_{i(t)}p[i(t)]e^{zt^*[i(t)]}.
\end{equation}
The summation over $i(t)$ is shorthand for a sum over all possible
sequences of states visited by $i(t)$ and integration over all time intervals
between transitions. We now consider an auxiliary ensemble, in which the same
trajectories $i(t)$ have a different weight $\hat p[i(t)]$. Using Jensen's
inequality, we can construct a bound on the CGF of the form
\begin{align}
  \ln\mean{e^{zt^*}}&=\ln\sum_{i(t)}\hat
  p[i(t)]\exp\left[\ln\frac{p[i(t)]}{\hat
  p[i(t)]}+zt^*[i(t)]\right]\nonumber\\
  &\geq \mean{\ln\frac{p[i(t)]}{\hat p[i(t)]}+zt^*[i(t)]}_\mathrm{aux},
  \label{eq:cgf_wtd_gen}
\end{align}
where the average with subscript `aux' is taken in the auxiliary
ensemble. 

In the auxiliary ensemble, we choose the weights $\hat p[i(t)]$ again according to the frequency of snippets of a long trajectory cut into pieces at every tick. But now, the long trajectory is generated with different rates
$\hat k_{ij}$ and accordingly a different stationary distribution $\hat
p_i$, traffic $\hat \traf_{ij}=\hat p_i\hat k_{ij}+\hat p_j \hat k_{ji}$, and current $\hat \jmath_{ij}=\hat p_i\hat k_{ij}-\hat p_j \hat k_{ji}$. The path weight $\hat p[i(t)]$ takes the same form as in
Eq.~\eqref{eq:pathweight}, but with the rates and initial distribution
replaced by the quantities for the auxiliary ensemble. We thus obtain
\begin{align}
  \ln\mean{e^{zt^*}}\geq  \mean{\ln\frac{p_{i(0)}(0)}{\hat p_{i(0)}(0)}}_\mathrm{aux}-\sum_i(w_i-\hat w_i)\mean{t_i}_\mathrm{aux}\nonumber\\+\sum_{ij}\mean{n_{ij}}_\mathrm{aux}\ln\frac{k_{ij}}{\hat
  k_{ij}}+z\mean{t^*}_\mathrm{aux}.
  \label{eq:cgf_wt_bnd}
\end{align}
The first term in Eq.~\eqref{eq:cgf_wt_bnd} is the Kullback-Leibler divergence $-D_\mathrm{KL}[\hat
p(0)\|p(0)]$ of the initial distributions, with $\hat p_i(0)$ being the distribution of the state $i$ immediately after a tick in the auxiliary ensemble. The average rate of ticks in the auxiliary ensemble is generally $\dN_\mathrm{aux}=\sum_{ij}b_{ij}\hat p_i\hat k_{ij}$. 

For the other terms in Eq.~\eqref{eq:cgf_wt_bnd}, consider again a very long trajectory generated with the
auxiliary transition rates, before cutting it into pieces. As argued before for the stationary ensemble, this trajectory has the length $t_\mathrm{obs}$ and the number of ticks $\dN_\mathrm{aux}t_\mathrm{obs}$, such that we get
$\mean{t^*}_\mathrm{aux}=1/\dN_\mathrm{aux}$. Likewise, the total time
$t_\mathrm{obs}\hat p_i$ spent in state $i$ and the total number of transitions $t_\mathrm{obs}\hat p_i\hat k_{ij}$ from
$i$ to $j$ are the sums all the observations of $t_i$ and $n_{ij}$,
respectively, over all the pieces of the long trajectory, and hence
$\mean{t_i}_\mathrm{aux}=\hat p_i/\dN_\mathrm{aux}$ and
$\mean{n_{ij}}_\mathrm{aux}=\hat p_i\hat k_{ij}/\dN_\mathrm{aux}$. 
We can then use any ansatz for $\hat k_{ij}$ with matching $\hat p_i$ to obtain
a bound on the CGF for the waiting time.

We now turn to ticks produced by the traffic of a single edge $(1,2)$. After a tick, the state of the system can only be $i(0)=1$ or
$i(0)=2$, depending on the state before the tick. From transitions in the stationary ensemble starting in state 2 and ending in state 1, we get $p_1(0)=\pstat_2 k_{21}/\dN$. For the other direction we get analogously $p_2(0)=\pstat_1 k_{12}/\dN$. The stationary tick rate
$\dN=\pstat_1 k_{12}+\pstat_2k_{21}$ serves as
normalisation. The initial distribution can also be expressed as $p_{1,2}(0)=(1\mp j\stat_{12}/\traf\stat_{12})/2$. Likewise, we obtain in the auxiliary ensemble $\hat p_{1,2}(0)=(1\mp \hat \jmath_{12}/\hat\traf_{12})/2$.

We use the same ansatz for the distribution $\hat p_i$, the traffic $\hat \traf_{ij}$, and the current $\hat \jmath_{ij}$ as for the bound on the large deviation function in Sec.~\ref{sec:ldf_traffic}. The rates of the auxiliary ensemble then follow as $\hat k_{ij}=(\hat\traf_{ij}+\hat \jmath_{ij})/(2\hat p_i)$. The rate of ticks is $\dN=\traf\stat_{12}$ in the stationary ensemble and $\dN_\mathrm{aux}=\hat\traf_{12}=\nu\traf\stat_{12}$ in the auxiliary ensemble. 
Plugging this ansatz into Eq.~\eqref{eq:cgf_wt_bnd}
and
maximising over the free parameters $\xi$ and $\nu$ to get the strongest possible bound, we obtain
\begin{equation}
  \ln\mean{e^{zt^*}}\geq\max_{\xi,\nu}\frac{1}{\nu}\left[\frac{z}{\langle\dot
      N\rangle}-B_\mathrm{t}^*(\bar\sigma,x,\xi,\nu)\right]
  \label{eq:cgf_wtd}
\end{equation}
with
\begin{align}
  B_\mathrm{t}^*(\bar\sigma,x,\xi,\nu)=\left(\bar\sigma-x\ln\frac{1+x}{1-x}\right)\frac{(\xi-1)^2}{4}+1-
    \nu-\nu\ln\nu\nonumber\\+(\nu+\xi x)\ln\frac{\nu+\xi x}{1+x}+(\nu-\xi
      x)\ln\frac{\nu-\xi x}{1-x}.
    \label{eq:Bdef_wtd}
\end{align}
For unknown $x$ we minimise the bound over $x$ to find the weakest possible
bound.

Note that the difference between $B_\mathrm{t}^*(\bar\sigma,x,\xi,\nu)$ and
$B_\mathrm{t}(\bar\sigma,x,\xi,\nu)$ of Eq.~\eqref{eq:Btraf} stems entirely from the
first term of Eq.~\eqref{eq:cgf_wt_bnd}. It can be attributed to the resetting
of the initial distribution after a tick (see, in particular, \cite{Monthus2021,Mori2022}), when we evaluate the fluctuations of uncorrelated samples of waiting times.

The coefficients for the quadratic expansion of $B_\mathrm{t}^*$ around $\nu=1$ and
$\xi=1$, analogous to Eq.~\eqref{eq:coeff_traffic}, read
\begin{equation}
\begin{split}
  \alpha &= \frac{2x^2}{1-x^2}+\frac{\bar\sigma}{2}-x\artanh(x), \\
  \beta&= -\frac{2x^2}{1-x^2}, \\ 
  \gamma &= \frac{1+x^2}{1-x^2} \, .
  \label{eq:coeff_wtd}
\end{split}
\end{equation}
This expansion is sufficient to perform the
maximisation over $\xi$ and $\nu$ in quadratic order in $z$, yielding
\begin{equation}
    \varepsilon^2=\dN^{2}\partial_z^2\left.\ln\mean{e^{zt^*}}\right|_{z=0}\geq\min_x\left[\frac{\alpha}{\alpha\gamma-\beta^2}\right] 
  \label{eq:localbound_wtd_gen}
\end{equation}
and finally
\begin{align}
  \varepsilon^2\geq\min_x\frac{\bar\sigma(1-x^2)+4x^2-2(1-x^2)x\artanh(x)}{\bar\sigma(1+x^2)+4x^2-2(1+x^2)x\artanh(x)}
  \label{eq:localbound_wtd}
\end{align}
as the bound on the fluctuations of the waiting time for the traffic.
For a parametric plot of this bound, as shown in Fig.~\ref{fig:allbounds}, we use Eq.~\eqref{eq:wtd_parametric} of Appendix~\ref{sec:parametric}. 

The same derivation as above applies to the waiting times between increments
of the flow of a single edge $(1,2)$. In that case, though, the first term of
Eq.~\eqref{eq:cgf_wt_bnd} vanishes, as the uniqueness of the state immediately
after a tick enforces $p_i(0)=\hat p_i(0)=\delta_{i2}$. As a consequence, we
get again Eq.~\eqref{eq:cgf_wtd}, but with $B_\mathrm{t}^*(\bar\sigma,x,\xi,\nu)$ replaced
by $B_\mathrm{f}(\bar\sigma,y,\xi,\nu)$, the same function that applies to the bound on
the large deviation function in Eq.~\eqref{eq:Bflow}. We thus get the same expansion coefficients, entering into Eq.~\eqref{eq:localbound_wtd_gen} in the same way as in the bound on the Fano factor of Eq.~\eqref{eq:localbound_general}. The bound on $\varepsilon^2$ is therefore the same as
the bound on $F$ in Eq.~\eqref{eq:localbound_flow}. This equality is in agreement with renewal property of the ticking process for this type of observable, by which $F=\varepsilon^2$. (Note, however, that this equality is no longer granted when ticks are produced by more than one edge, as will be discussed in Sec.~\ref{sec:multiedge_wtd}.)

\section{Optimal networks}
\label{sec:optimal}

We now turn again to the analysis of concrete networks, and seek to saturate the bounds on precision.
We use the exact expressions for the precision of Sec.~\ref{sec:illustration} and take cues from the numerics presented there. The latter have shown that optimal networks are unicyclic with a single visible edge $(M,1)$. 
Upon closer inspection, we find that for these optimal networks, the stationary probability is equal in the two states connected by the visible edge, $\pstat_M=\pstat_1=:p_A$, and equal for all other states $i\notin\{M,1\}$, $\pstat_i=:p_B$. Moreover, we find that the flow $n\stat_{ij}=\pstat_i k_{ij}$ is equal for all forward transitions and equal for all backward transitions, except for the visible edge (but including the two neighbouring edges).

We take networks of this form as candidates for a possible saturation of our bounds. For a convenient parameterisation, we fix the cycle current to $n\stat_{i,i+1}-n\stat_{i+1,i}=1$, other values can be obtained by scaling all rates equally, without changing the dimensionless precision and $\bar\sigma$. We use the affinities $A:=\ln(n\stat_{M1}/n\stat_{1M})$ of the visible edge $B=\ln(n\stat_{i,i+1}/n\stat_{i+1,i})$ of all other edges ($i\neq M$) along with $p_A$ as parameters. The probability $p_B$ is fixed through $2p_A+(M-2)p_B=1$.
The transition rates can be expressed in terms of these parameters as
\begin{align}
  k_{M1}&=\frac{1}{p_A(1-e^{-A})}, &k_{1M}&=\frac{1}{p_A(e^{A}-1)}\label{eq:optrate1},\\
  k_{12}&=\frac{1}{p_A(1-e^{-B})}, &k_{M,M-1}&=\frac{1}{p_A(e^{B}-1)}\label{eq:optrate2},\\
  k_+&=\frac{1}{p_B(1-e^{-B})}, &k_-&=\frac{1}{p_B(e^{B}-1)}\label{eq:optrate3},
\end{align}
where $k_\pm$ are the forward/backward transition rates for the bulk of edges. 

The tilted matrix \eqref{eq:TiltedMatrix} for our candidates of networks reads as follows:
\begin{widetext}
\begin{equation}
    \label{eq:TiltedAnalyticFano}
    \tilde{\boldsymbol{L}}(s) = 
    \begin{pmatrix}
    -k_{12} - k_{1M} & k_{12} & 0 & \cdots & 0 & k_{1M} e^{b s} \\
    k_- & - k_+ - k_+ & k_+ & \cdots & 0 & 0\\
    0 & k_- & -k_- - k_+ & k_+ & \cdots & 0\\
    0 & 0 & k_- & -k_- -k_+ & \cdots & 0\\
    \vdots & & \ddots & & & \vdots  \\
    k_{M1} e^{s} & 0 & 0 & \cdots & k_{M, M-1} & -k_{M1} -k_{M, M-1} 
    \end{pmatrix}.
\end{equation}
\end{widetext}
We use $b\coloneqq b_{M1}$ as a parameter whose value determines the kind of observable we look at: $b=0$ for the flow from $M$ to $1$, and $b=1$ for the traffic through $(M,1)$.

The entropy production~\eqref{eq:sigmatot} in this class of networks is simply $\sigma=A+(M-1)B$. With the rate of ticks $\dN=p_A(k_{M1}+b\, k_{1M})$ we then get
\begin{equation}
    \bar\sigma=\frac{e^A-1}{e^A+b}[A+(M-1)B].
    \label{eq:sigmabar_uc}
\end{equation}
Keeping $\bar\sigma$ fixed thus reduces the optimisation to a two-dimensional one.

Focusing on the sub-matrix obtained by discarding the first and last row of \eqref{eq:TiltedAnalyticFano}, we end up with a tridiagonal structure that can be used to simplify the eigenvalue problem presented in \eqref{eq:ev}. Indeed, the now so-called ``bulk'' equation for the eigenvalue problem for a general row $i \in \left[ 2, M-1 \right]$ reads
\begin{equation}
    k_- r_{i-1}(s) - (k_- + k_+) r_i(s) + k_+ r_{i+1}(s) = \Psi(s) r_i(s) \ .
    \label{eq:BulkEquation}
\end{equation}
In principle, one could solve this recursion relation exactly, determining $r_i$ for all bulk states for just two (yet to be determined) initial conditions $r_1$ and $r_2$, reducing the $M$-dimensional algebraic problem to just two dimensions. 

For the calculation of the Fano factor, it is sufficient to expand the eigenvalue to second order and the eigenvector to first order in $s$,
\begin{align}
    \label{eq:EigvalueExpansion}
    \Psi(s) &= 0 + s \dN + \frac{1}{2}\dN F s^2+ \mathcal{O}(s^3) \\
    \label{eq:EigvectorExpansion}
    r_i(s) &= 1 + s \tilde{r}_i + \mathcal{O}(s^2).
\end{align}
Since the tilted matrix \eqref{eq:TiltedAnalyticFano} is stochastic for $s=0$, its dominant eigenvalue is then $0$ and its dominant right eigenvector is $\boldsymbol{1}$. Taking further cues from the numerics for optimal networks, we use the linear ansatz
\begin{equation}
    \label{eq:EducatedAnsatz}
    \tilde{r}_i = q + i m,
\end{equation}
with yet to be determined parameters $q$ and $m$.
Plugging everything back in \eqref{eq:BulkEquation}, we obtain at first order in $s$ the following equation for $m$
\begin{equation}
    \label{eq:mEquation}
    k_-\, (i-1)m - (k_- + k_+) i m + k_+\, (i+1) m = \dN \ ,
\end{equation}
which is solved by
\begin{equation}
    \label{eq:mSolution}
    m = \frac{\dN}{k_+ - k_-} =p_B\dN .
\end{equation}
The coefficient $q$ is fixed by applying the constraint~\eqref{eq:ev_norm} on the dominant left and right eigenvectors.
As the leading order of $l_i$ is the stationary distribution $\pstat_i$, this constraint simplifies at first order in $s$ to
\begin{equation}
    \label{eq:NormalisationConstraint2}
    \sum_{i=1}^M \pstat_i \tilde{r}_i = 0 \ .
\end{equation}
Plugging the specific form of the stationary distribution given by $p_A$ and $p_B$ into \eqref{eq:NormalisationConstraint2}, we obtain an equation which is solved by
\begin{equation}
    \label{eq:qSolution}
    q = - \frac{1}{2}\dN (M+1) p_B \ .
\end{equation}

The linear ansatz~\eqref{eq:EducatedAnsatz} for $\tilde r_i$ also needs to be consistent with the first and last row of the matrix Eq.~\eqref{eq:TiltedAnalyticFano} in the eigenvalue problem. From the first row, we get
\begin{equation}
\label{eq:Constraints1}
(-k_{12}-k_{1M}) r_1 + k_{12} r_2 + k_{1M} e^{b s} r_M = \Psi(s) r_1,
\end{equation}
or, to linear order in $s$,
\begin{equation}
\label{eq:Constraints2}
(-k_{12}-k_{1M}) \tilde r_1 + k_{12} \tilde r_2 + k_{1M} (\tilde r_M +b)= \dN.
\end{equation}
Plugging in the ansatz~\eqref{eq:EducatedAnsatz} and solving for $p_B$ we get
\begin{equation}
    p_B=\frac{(e^A-b)}{(e^A+b)[(M-1)\coth(A/2)+\coth(B/2)]}
    \label{eq:pb_const}
\end{equation}
as a further constraint, thus reducing the dimensionality of the optimisation problem to just one variable. Repeating this calculation for the last row of Eq.~\eqref{eq:TiltedAnalyticFano} yields the same expression for $p_B$, which shows that the linear ansatz is consistent with this choice for $p_B$ (and accordingly $p_A$).

By making use of the first-order analytic form of $\tilde{\boldsymbol{r}}$ in \eqref{eq:EducatedAnsatz}, we can calculate the Fano factor from the second order of the eigenvalue problem~\eqref{eq:EigvalueExpansion}. This reads
\begin{align}
    F &= \frac{1}{\dN}\left[\boldsymbol{p}\stat\tilde{\boldsymbol{L}}''(0)\boldsymbol{1}+2\boldsymbol{p}\stat\tilde{\boldsymbol{L}}'(0)\tilde{\boldsymbol{r}}\right] \nonumber \\
    &= 1+\frac{2p_A}{\dN}( b \, k_{1M} \tilde{r}_M + k_{M1} \tilde{r}_1  ) \ ,
    \label{eq:AnalyticFanoFactor}
\end{align}
where $'$ denotes derivatives for $s$. We plug in $\tilde{r}_1$ and $\tilde{r}_M$ as given by \eqref{eq:EducatedAnsatz} with $m$ and $q$ from, respectively, Eq.\ \eqref{eq:mSolution} and Eq.\ \eqref{eq:qSolution}. This finally simplifies to
\begin{equation}
    F=1-(M-1)p_B\frac{e^A-b}{e^A-1}.
\end{equation}
Now we can express $B$ through $A$ and $\bar\sigma$ using Eq.~\eqref{eq:sigmabar_uc} and substitute $A=2\artanh(x)$ for the traffic ($b=1$) and $A=-\ln(1-y)$ for the flow ($b=0$). Taking the limit $M\to\infty$ and optimising over $x$ and $y$, respectively, then recovers the bounds~\eqref{eq:localbound_fano} and \eqref{eq:localbound_flow}. This shows that the optimal networks constructed in this way can indeed saturate these bounds.

The uncertainty of the waiting time between ticks can also be minimised within the class of unicyclic networks parameterised by the rates~\eqref{eq:optrate1}-\eqref{eq:optrate3}. For the flow of a single edge, the Fano factor and the uncertainty of waiting times are the same (see Sec.~\ref{sec:wtd} above), such that the optimal precision is obtained in the same network, for both measures of precision. We therefore focus on the traffic only, setting $b=1$. Taking the limit $s\to-\infty$ in the matrix~\eqref{eq:TiltedMatrix}, we obtain the matrix $\boldsymbol{S}$. We now define for the expression in Eq.~\eqref{eq:cv_exact} $\boldsymbol{r}\coloneqq\boldsymbol{S}^{-1}\boldsymbol{1}$. It plays a similar role to the right eigenvector above, hence the same symbol. Namely, it turns out that in optimal networks it takes the linear form $r_i=q+im$. The bulk part of the equation $\boldsymbol{Sr}=\boldsymbol{1}$ fixes $m=p_B$, while the first and last row determine $q$ and
\begin{equation}
    p_A=\frac{e^A[e^{A+B}+(M-2)(e^B-1)]-1}{(M-2)(e^{2A}-e^B)+M(1-e^{2A+B})}.
\end{equation}
From Eq.~\eqref{eq:cv_exact} we then obtain
\begin{equation}
    \varepsilon^2=-\dN[2q+(M+1)p_B]-1.
\end{equation}
Using Eq.~\eqref{eq:sigmabar_uc}, the substitution $A=2\artanh(x)$, and the limit $M\to\infty$ then yields a saturation of the bound~\eqref{eq:localbound_wtd}.

\section{Generalisation to several visible edges}
\label{sec:multiedge}

So far, the bounds on precision we have derived apply to the case of ticks stemming from a single edge of the Markov network. We now turn to the case where several visible edges contribute to the ticking process, without distinguishing which of these edges produced a tick.  We write $\mathcal{T}=\{(i,j)|i<j\land
b_{ij}=b_{ji}=1\}$ for the set of edges contributing via their traffic (labeled only
in one direction) and $\mathcal{F}=\{(i,j)|b_{ij}=1\land b_{ji}=0\}$ for the set of directed edges contributing via their flow. For notational convenience, we refer to elements of either of these sets by a single letter $a$. We will see that in most cases mixing the observations from several visible edges does not improve the precision at fixed entropic cost. The only exception is the case of the waiting time fluctuations of flow-like observables.

\subsection{Bounds on the Fano factor}
\label{sec:multiedge_fano}

We start by considering the fluctuations of $N(t)$ for traffic-like observables. There, we can write $N(t)=t\sum_{a}\traf_a$, where the sum runs over all $a\in\mathcal{T}$. Scaling by the overall rate $\dN$ leads to $\nu=N(t)/\dN=\sum_a c_a \nu_a$, where we identify normalized weights $c_a \coloneqq \traf\stat_a/\dN$ by which the individual visible edges contribute to the overall tick rate, and the individual scaled traffic observables $\nu_a \coloneqq \traf_a/\traf\stat_a$. The joint distribution of fluctuations of the traffic observables is captured in the long-time limit by the large deviation function $I(\{\nu_a\})$. A bound on this function can be found in a similar way as in Sec.~\ref{sec:ldf_traffic}, but now using an ansatz that keeps the traffic of not just one but all visible edges fixed. Scaling by $\dN$, we obtain
\begin{equation}
  \bar I(\{\nu_a\})\leq\min_\xi\sum_a c_a B_\mathrm{t}(\bar\sigma,x_a,\xi,\nu_a)
\end{equation}
with $x_a \coloneqq j\stat_a/\traf\stat_a$, which at first we assume to be a given set of parameters (along with $c_a$). The large deviations for the overall tick rate then follow through contraction as
\begin{equation}
  \bar I(\nu)\leq\min_{\{\nu_a\}|\sum_ac_a\nu_a=\nu}\min_\xi\sum_a c_a B_\mathrm{t}(\bar\sigma,x_a,\xi,\nu_a).
\end{equation}
The constrained minimisation over all $\nu_a$ with $a\in\mathcal{T}$ can be turned into an unconstrained one after a Legendre transform of $\bar{I}(\nu)$. For the SCGF we then find the bound
\begin{align}
  \bar\Psi(s)& = \max_\nu[\nu s-\bar I(\nu)]\nonumber\\
  &\geq\max_\xi\max_{\{\nu_a\}}\sum_ac_a[s\nu_a-B_\mathrm{t}(\bar\sigma,x_a,\xi,\nu_a)].
\end{align}
In principle, it is possible to carry out the minimisation over
$\nu_a$ for each term individually, leading to the optimal
$\nu_a(\xi)=\sqrt{(1-x_a^2)\exp(2s)+(\xi x_a)^2}$. However, the remaining
optimisation over $\xi$ remains analytically non-tractable. 

To make further
analytical progress, we focus on typical fluctuations $\nu\approx 1$, where
we can also expect $\xi\approx 1$ and all $\nu_a\approx 1$. Here we can make
use of the expansion of the function $B(\bar\sigma,x_a,\xi,\nu_a)$ up to quadratic order in $\xi$ and $\nu_a$, as derived in
Eq.~\eqref{eq:quadform}. This yields
\begin{align}
   \bar\Psi(s)\geq \max_{\xi,\{\nu_a\}}\sum_a
   c_a\Big[s\nu_a&-\frac{1}{2}\alpha_a(\xi-1)^2-\beta_a(\xi-1)(\nu_a-1)\nonumber\\
   &-\frac{1}{2}\gamma_a(\nu_a-1)^2+\mathrm{h.o.t.}\Big],
   \label{eq:quadraticform_multi}
 \end{align}
 with $\alpha_a$, $\beta_a$, $\gamma_a$ defined as in
 Eq.~\eqref{eq:coeff_traffic}, but with $x$ replaced by $x_a$. Higher order terms (h.o.t.)~are not relevant for bounding the variance of fluctuations.
 The optimal values of $\nu_a$ and $\xi$ follow as
 \begin{equation}
   \xi^*-1=C s+\mathcal{O}(s^2)
 \end{equation}
 and
 \begin{equation}
   \nu_a^*-1=\frac{1-\beta_aC}{\gamma_a} s+\mathcal{O}(s^2)
 \end{equation}
 with
 \begin{equation}
   C \coloneqq \frac{\sum_a c_ax_a^2}{\bar\sigma/2+\sum_ac_a[x_a^2-x_a\artanh(x_a)]}.
 \end{equation}
 Plugging these into Eq.~\eqref{eq:quadraticform_multi} and collecting terms yields the bound on
 the SCGF $\bar\Psi(s)\geq s+F_B s^2/2+\mathcal{O}(s^3)$, where we identify as bound on the Fano factor 
 \begin{equation}
   F=\partial_s^2\bar\Psi(s)|_{s=0}\geq F_B\coloneqq
   \sum_a
   \frac{c_a}{\gamma_a}+\frac{\left(\sum_ac_a\beta_a/\gamma_a\right)^2}{\sum_ac_a(\alpha_a-\beta_a^2/\gamma_a)}.
   \label{eq:fano_bound_general}
 \end{equation}
With the functional form of Eq.~\eqref{eq:coeff_traffic} for
$\alpha_a,\beta_a,\gamma_a$, this bound becomes
  \begin{equation}
F_B=-\sum_a
   c_ax_a^2+\frac{\left(\sum_ac_a x_a^2\right)^2}{\bar\sigma/2-\sum_ac_a[x_a\artanh(x_a)+x_a^2]}.
   \label{eq:trafficbound_detailed}
   \end{equation}
Note that, since the sum of the individual entropy production rates $\sigma_a$ of all visible edges cannot be greater than the total entropy production in the denominator, we have that $\sum_a
c_ax_a\artanh(x_a)=\sum_a\sigma_a/(2\langle\dot N\rangle)\leq\bar\sigma/2$ . 

In the form of Eq.~\eqref{eq:trafficbound_detailed}, the bound is only useful if $x_a$ and $c_a$ is known for all the visible edges. As this is typically not the case, we 
replace $z_a=x_a^2$ and use the convexity of
$\sqrt{z}\artanh\sqrt{z}$ (straightforward to check) to get the Jensen's
inequality $\sum_ac_a\sqrt{z_a}\artanh\sqrt{z_a}\geq x\artanh(x)$. Here, we have
generalised the definition of $x$ (defined previously for the case of a single
visible edge) to $x \coloneqq \sqrt{\sum_ac_ax_a^2}$.
The parameter $x$ is still unknown, such that we need to take the
loosest possible bound. Through the minimisation over $x$, we recover
Eq.~\eqref{eq:localbound_fano} derived previously for a single
visible edge.

Note that in Jensen's inequality, equality holds either if all edges
contributing to the observed traffic have equal $x_a$ (and hence equal
affinity), or if all but one of the visible edges have a non-zero $c_a$
(and hence non-zero traffic $\traf\stat_a$). Edges with zero traffic can be
removed from the network without harm, such that in the latter case we are
back to the single visible edge case. 

To obtain a bound for the fluctuations of pure flow-like observables (with the set $\mathcal{F}$ containing the visible edges), we can follow the same steps as above, with $B_\mathrm{t}$ replaced by $B_\mathrm{f}$.
Plugging the coefficients \eqref{eq:coeff_flow} into Eq.~\eqref{eq:fano_bound_general} then leads to 
\begin{equation}
  F\geq\frac{1}{2}-\frac{1}{2}\sum_a c_a z_a+\frac{\left(\sum_a
      c_az_a\right)^2}{\frac{\bar\sigma}{2}+\sum_a c_a\frac{2z_a}{1+z_a}\left(\frac{1}{2}\ln\frac{1-z_a}{1+z_a}+z_a\right)}
\end{equation}
with $c_a=n\stat_a/\langle\dot N\rangle$, $y_a=j\stat_a/n\stat_a$ and the
substitution $z_a=y_a/(2-y_a)$. Summation is implied over all $a\in\mathcal{F}$. Using Jensen's inequality with the convex
function $\frac{2z}{1+z}\left(\frac{1}{2}\ln\frac{1-z}{1+z}+z\right)$ (in the
entire domain $-1<z<1$), setting $z=\sum_ac_a z_a$, re-substituting
$y=2z/(1+z)$, and minimising over $y$, we recover
Eq.~\eqref{eq:localbound_flow} as a bound that is also valid for flow-like
observables involving several edges. We show in Appendix \ref{sec:mixed_obs} that this bound is also valid for the generic case of a mixed-type observable where both $\mathcal{T}$ and $\mathcal{F}$ are non-empty sets.

\subsection{Bounds on waiting time fluctuations}
\label{sec:multiedge_wtd}

We now turn to the fluctuations of waiting times, generalising Sec.~\ref{sec:wtd} to the case of several visible edges.
The derivation there, up to and including the paragraph after Eq.~\eqref{eq:cgf_wt_bnd}, is valid for general counting observables, including the case of several visible edges. We now need to generalise the form of the distribution $p_i(0)$ immediately after a tick, which is no longer limited to the states~$1$ and~$2$.
Instead, we get in the stationary state
\begin{equation}
  p_i(0)=\sum_{j}b_{ji}\pstat_jk_{ji}/\langle\dot
  N\rangle=\sum_jb_{ji}\frac{\traf\stat_{ij}-j\stat_{ij}}{2\langle\dot N\rangle}
\end{equation}
and likewise in the auxiliary ensemble
\begin{equation}
  \hat p_i(0)=\sum_jb_{ji}\frac{\hat\traf_{ij}-\hat\jmath_{ij}}{2\langle\dot N\rangle_\mathrm{aux}}.
\end{equation}
Using the log-sum inequality, the Kullback-Leibler divergence of
Eq.~\eqref{eq:cgf_wt_bnd} can be bounded by
\begin{align}
  -& D_\mathrm{KL}[\hat p(0)\|p(0)]\nonumber\\
  &\geq\ln\frac{\langle\dot
    N\rangle_\mathrm{aux}}{\langle\dot N\rangle}+\frac{1}{2
    \langle\dot
    N\rangle_\mathrm{aux}}\sum_{ij}b_{ij}(\hat\traf_{ij}-\hat\jmath_{ij})\ln\frac{\traf\stat_{ij}-j\stat_{ij}}{\hat\traf_{ij}-\hat\jmath_{ij}}.
\end{align}
Using the ansatz of Sec.~\ref{sec:multiedge_fano} for the traffic and flow of the auxiliary
ensemble, we obtain
\begin{align}
  &\ln\mean{e^{zt^*}}\nonumber\\
  &\geq
    \max_{\xi,\{\nu_a\}}\left(\frac{1}{\nu}\sum_ac_a\left[\frac{z}{\langle\dot
        N\rangle}-B_a(\bar\sigma,\xi,\nu_a)-\nu_a\ln\nu_a\right]+\ln\nu \right),
        \label{eq:cgf_multiedge}
\end{align}
with summation over $a\in\mathcal{T}\cup\mathcal{F}$,
where $B_a(\bar\sigma,\xi,\nu_a)$ is either $B_\mathrm{t}^*(\bar\sigma,x_a,\xi,\nu_a)$ for $a\in\mathcal{T}$ or $B_\mathrm{f}(\bar\sigma,y_a,\xi,\nu_a)$
for $a\in\mathcal{F}$, and where $\nu=\langle\dot
N\rangle_\mathrm{aux}/\langle\dot N\rangle=\sum_a c_a\nu_a$. The
expression to be maximised can be expanded in quadratic order around
$\xi=1$ and $\nu_a=1$, as  shown in Appendix~\ref{sec:hessian}. This finally yields as bound on the uncertainty of the waiting time
\begin{equation}
  \varepsilon^2\geq C/(1-C)
  \label{eq:epsilon_C}
\end{equation}
with
\begin{equation}
  C \coloneqq \sum_a\frac{c_a}{\gamma_a+1}+\frac{\left(\sum_a\frac{c_a\beta_a}{\gamma_a+1}\right)^2}{\sum_ac_a\left(\alpha_a-\frac{\beta_a^2}{\gamma_a+1}\right)}.
  \label{eq:wtd_general}
\end{equation}

For pure traffic-like observables, the coefficients
$\alpha_a,\beta_a,\gamma_a$ depend on $x_a$ like
Eqs.~\eqref{eq:coeff_wtd}. Eq.~\eqref{eq:wtd_general} then becomes
\begin{equation}
  C=\sum_a\frac{2c_a}{1-x_a^2}+\frac{\left(\sum_ac_ax_a^2\right)^2}{\bar\sigma/2-\sum_ac_ax_a\artanh(x_a)+2\sum_ac_ax_a^2}.
\end{equation}
Using the substitution $z_a=x_a^2$, the convexity of all relevant
terms in $z_a$, and the monotonicity of Eq.~\eqref{eq:epsilon_C} in
$C$, and setting $x=\sqrt{\sum_ac_ax_a^2}$, we finally recover the
bound~\eqref{eq:localbound_wtd} that we have derived before for a single visible edge.

\subsection{Phase transition for flow-type observables}
\label{sec:phasetrans}

\begin{figure}
  \centering
  \includegraphics{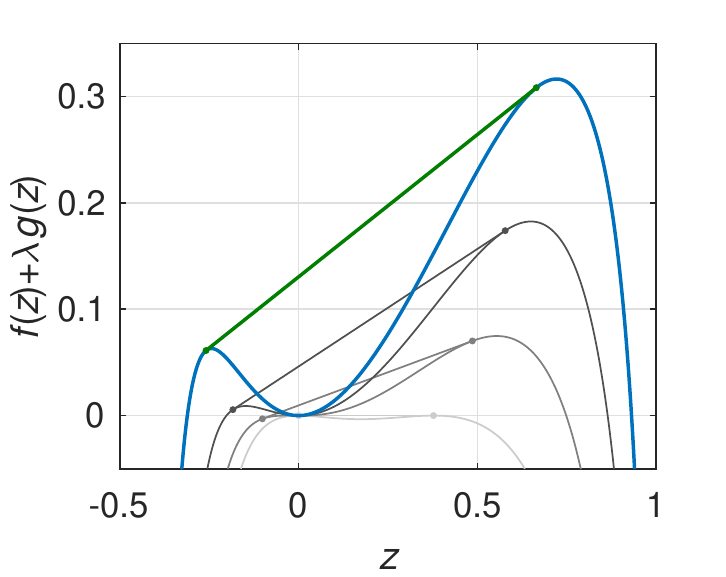}
  \caption{Plot of the function $f(z)$ of Eq.~\eqref{eq:nonconvex}
    (blue), along with its convex hull (green). For small values of
    $\bar\sigma$ (discussed in Appendix~\ref{sec:smallsigma}), we add the negative function $\lambda g(z)$  with
    $\lambda>0$ up to $\lambda_\mathrm{max}$ (from dark to light grey)}
  \label{fig:convexhull}
\end{figure}

The general bound on waiting time fluctuations, Eq.~\eqref{eq:epsilon_C} and \eqref{eq:wtd_general}, also holds for pure flow-type observables. We use the coefficients of Eq.~\eqref{eq:coeff_flow} and substitute $z_a=y_a/(3-2y_a)$ (note the domain
$-1/2\leq z_a\leq 1$). The quantity $C$ of Eq.~\eqref{eq:wtd_general} then becomes
\begin{equation}
  C=\frac{1}{3}-\frac{1}{3}\sum_a c_a z_a+\frac{\left(\sum_a c_a
      z_a\right)^2}{\bar\sigma/2+\sum_a c_a f(z_a)}
  \label{eq:localbound_flow_wtd_detailed}
\end{equation}
with the function
\begin{equation}
  f(z)=\frac{3z}{1+2z}\left(\frac{1}{2}\ln\frac{1-z}{1+2z}+2z\right).
  \label{eq:nonconvex}
\end{equation}
Remarkably, this function is not convex, as shown in
Fig.~\ref{fig:convexhull}. This means that we cannot use a simple convexity
argument as in the bounds derived so far to show that the
bound for several visible edges is the same as for a single
one. 

By construction, in the case of a single visible edge (such that $\varepsilon^2=F$ for the flow), the bound~\eqref{eq:localbound_flow_wtd_detailed} reverts to the bound~\eqref{eq:localbound_flow}. However, for two visible edges with two different values $z_a$, we can obtain a bound on $\varepsilon^2$ that is smaller than that, at the same entropy production rate $\bar\sigma$. The condition is that the optimal value for $y$, as given by Eq.~\eqref{eq:flow_parametric}, is such that $z=y/(3-2y)$ falls into the non-convex region of $f(z)$. This is the case for all $\bar\sigma$ below a critical value $\bar\sigma_c\simeq 7.1763$. Then, it is possible to choose $z_1$ and $z_2$, from the non-convex region such that $\sum_ac_az_a=z$ and $\sum_ac_af(z_a)>f(z)$, leading to a smaller bound on $\varepsilon^2$. This is similar to the effect of a phase transition, where the overall free energy can be lowered by separating a system into two components with different free energy densities. For
values of $\bar\sigma$ larger than $\bar\sigma_c$, the best precision can be
obtained by observing ticks stemming from a single edge, for smaller
values of $\bar\sigma$ an improvement of the precision can be obtained
by combining the signal of two edges with distinct affinities.

We can find a lower bound on $C$ by replacing $f(z)$ by its convex hull $f_c(z)\geq f(z)$, as shown in
Fig.~\ref{fig:convexhull}. We can then apply Jensen's inequality
to obtain the bound $\varepsilon^2\geq C'/(1-C')$ with
\begin{equation}
  C'=\min_z\left[\frac{1-z}{3}+\frac{z^2}{\bar\sigma/2+f_c(z)}\right].
  \label{eq:localbound_flow_wtd}
\end{equation}
The non-convex region is $\hat z_1<z<\hat z_2$, determined numerically as
$\hat z_1\simeq -0.2589$ and $\hat z_2\simeq0.6646$. This region corresponds to $-1.6102\lesssim y\lesssim 0.8560$.

Thanks to the linearity of the convex hull between $\hat z_1$ and $\hat z_2$, the
minimisation of Eq.~\eqref{eq:localbound_flow_wtd} can be carried out
analytically, yielding the optimal value
\begin{equation}
  z^*=\frac{\bar\sigma/2+f_0}{\varDelta} \left( -1+\sqrt{3/(3-\varDelta)} \right)
  \label{eq:zopt}
\end{equation}
with the constants
\begin{equation}
\begin{split}
  f_0 &\coloneqq \frac{f(\hat z_1)\hat z_2-f(\hat z_2)\hat z_1}{\hat z_2-\hat z_1} \\ 
  \varDelta &\coloneqq \frac{f(\hat z_2)-f(\hat z_1)}{\hat z_2-\hat z_1}
\end{split}
\end{equation}
and the function value $f_c(z^*)=f_0+z^*\varDelta$.

While Eq.~\eqref{eq:localbound_flow_wtd} provides a bound that is valid for
all $\bar\sigma$, it cannot be tight for small $\bar\sigma$. The coefficients
$c_a$ and $z_a$ corresponding to a point on the convex hull may be
incompatible with the requirement that the entropy production of all the observed links cannot be
more than the total entropy production. In Appendix~\ref{sec:smallsigma}, we derive a refined bound that is relevant for $\bar\sigma\lesssim1.5974$. Moreover, in Appendix~\ref{sec:mixed_obs}, we show that this bound also holds for mixed flow- and traffic-type observables.

In Fig.~\ref{fig:allbounds}, we show the bound~\eqref{eq:localbound_flow_wtd} with this refinement. 
It turns out that one can reduce
the entropic cost by up to 31\,\% compared to the case of a single
visible edge for uncertainties slightly below $1/2$. 
Using networks similar to the ones discussed in Sec.~\ref{sec:optimal} but with two visible edges (oppositely directed, to match the signs and values of $\hat z_{1,2}$), we see numerically that it is indeed possible to ``break'' the bound~\eqref{eq:localbound_flow} that holds for a single visible edge. The full analysis of the form of optimal networks in this case is left for future work. We just note in passing that for unicyclic networks, $x_a$ and $c_a$ are related through the constant current. Hence, it is only possible to change the weights given to $\hat z_1$ and $\hat z_2$ (as necessary for a saturation of the bound) by having much more than just two visible edges.

\section{Conclusions}
\label{sec:conclusions}

In this article, we have discussed the problem of inferring the rate of entropy production from the statistics of general counting observables that record ``ticks'', triggered by some internal changes in a Markovian system. We have analytically derived bounds that relate the variance of symmetric (traffic-like) and asymmetric (flow-like) observables to the rate of entropy dissipated by the system, hereby extending the well-known TUR valid for time-antisymmetric observables. We have derived these bounds by opportunely minimising the level 2.5 large deviations and only then assessing typical fluctuations. Furthermore, we have analytically obtained bounds that relate the rate of entropy production with the waiting time between ticks. In deriving these results we made use of a path-reweighting technique, similar to the level 2.5, applied on the cumulant generating function of the waiting time. The bounds are tight in the sense that they can be saturated, as we have shown explicitly for optimal unicyclic networks. All bounds were first derived by considering transitions stemming from a single visible edge of the Markov network and then extended to multiple visible edges. Interestingly for the latter, in the case of waiting time fluctuations of asymmetric observables, a phase transition to optimal configurations arises where higher precision can be achieved by combining the signals from different edges.

Our results apply to the vast class of systems that can, at some microscopic scale, be modeled as a Markov jump process. Just like for the TUR, this encompasses systems with one or several continuous degrees of freedom undergoing overdamped Brownian motion~\cite{dech18}, which can be arbitrarily finely discretised. However, there needs to be a clear separation between the states connected by a visible transition, in a way that is suitable for a Markovian description (see Ref.~\cite{hart21a} for an important caveat). Otherwise, frequent re-crossings of the visible edge will spoil the precision of the recorded signal. We leave it to future work to study bounds on precision in a fully continuous system, where discrete events are detected through a ``milestoning'' that suppresses fast re-crossings~\cite{fara04,hart21}.

The large deviation framework we have presented here is sufficiently versatile to generalise and refine bounds on precision. 
For simplicity, we have focused here on a definition of the Fano factor~\eqref{eq:fanodef} employing the long-time limit. However, it is straightforward to show that the same bounds apply to the Fano factor evaluated at any finite time, in analogy to the TUR~\cite{piet17,horo17,mani18}. Moreover, it should be possible to bound large deviation functions more carefully, taking into account the affinity and topology of cycles~\cite{bara15a,piet16a}. This will likely show that the precision calculated for the networks of Sec.~\ref{sec:optimal}, before taking the limit $M\to\infty$, is in fact optimal over the whole class of networks with a fixed, finite number of states $M$.

In our work we have focused on the minimal setting where ticks are only distinguishable by their point in time. While the resulting bounds have the broadest applicability, it may be possible to derive, in a similar manner, stronger bounds on the entropy production in more specific settings with more detailed information. In particular, an extension of the formalism of Sec.~\ref{sec:wtd} for waiting times between two alternating types of ticks may yield an analytical form and a proof of the bound obtained numerically in Ref.~\cite{skin21}.

An almost periodic sequence of ticks is an indicator for coherent oscillations in a system. In an underdamped oscillator with inertia, such oscillations can emerge in equilibrium, creating a precise sequence of ticks at zero entropic cost~\cite{piet22}. Hence, just like the TUR, the bounds derived here cannot be valid for underdamped Brownian dynamics. Emulating such oscillating behaviour in an overdamped system, as relevant for biological systems on the cellular scale, requires a minimal strength of driving~\cite{cao15}. For a universal quantification of coherence, this trade-off has been conjectured in Ref.~\cite{bara17} and recently proven~\cite{kolc23}. Yet, a proof for a relation involving the entropy production rate~\cite{ober22} is still outstanding. Our results could shed light on this problem, from the perspective of the number of oscillations as a counting observable.

\begin{acknowledgments}

The authors thank K.\ S.\ Olsen and J.\ Meibohm for insightful discussions and D.\ M.\ Busiello for feedback on a first complete draft of the manuscript. The authors also acknowledge the Nordita program ``Are There Universal Laws in Non-Equilibrium Statistical Physics?''~(May 2022), where this collaboration started. F.\ C.\ acknowledges support from the Nordita Fellowship program. This work was partially
supported by the Swedish Research Council Grant No. 638-
2013-9243.
\end{acknowledgments}

\appendix

\section{Parametric forms of the bounds and asymptotics}
\label{sec:parametric}

\begin{table}
    \centering
    \begin{tabular}{l||l|l}
    bound & small $\bar\sigma$ & large $\bar\sigma$\\\hline
    $F$ traffic & $F\geq 1-\bar\sigma/8$ & $\bar\sigma\gtrsim 2/F-2\ln F-1+3\ln2$\\
    $F$ flow & $F\geq 1/2-\bar\sigma/32$ & $\bar\sigma\gtrsim 2/F-2\ln F-1+\ln2$\\
    $\varepsilon^2$ traffic & $\varepsilon^2\geq1-(3-2\sqrt{2})\bar\sigma$ & $\bar\sigma\gtrsim 2/\varepsilon^2-2\ln \varepsilon^2-1+2\ln2$\\
    $\varepsilon^2$ flow & $\varepsilon^2\gtrsim 1/2-\bar\sigma/21.97$ & $\bar\sigma\gtrsim 2/\varepsilon^2-2\ln \varepsilon^2-1+\ln2$
    \end{tabular}
    \caption{Expansions of our four bounds for small and large~$\bar\sigma$.}
    \label{tab:asymptotics}
\end{table}

We write the function to be optimised for each of the bounds \eqref{eq:localbound_fano}, \eqref{eq:localbound_flow}, and \eqref{eq:localbound_wtd} as $f(x,\bar\sigma)$ (or $y$ instead of $x$ for the flow). It is possible to solve the stationarity condition $\partial_xf(\bar\sigma,x)=0$ for $\bar\sigma$, which then yields $f(x,\bar\sigma(x))$ and $\bar\sigma(x)$ as a parametric form of the bounds shown in Fig.~\ref{fig:allbounds}. For the Fano factor of the traffic, Eq.~\eqref{eq:localbound_fano}, we obtain
\begin{equation}
    \bar\sigma(x)=2x\artanh(x)+x\sqrt{2x\artanh(x)+\frac{2x^2}{1-x^2}},
\end{equation}
for the Fano factor of the flow, eq.~\eqref{eq:localbound_flow},
\begin{equation}
  \bar\sigma(y)=-y\ln(1-y)+2y+y\sqrt{2\frac{2-y}{1-y}-2\ln(1-y)},
  \label{eq:flow_parametric}
\end{equation}
and for the waiting time fluctuations of the traffic, Eq.~\eqref{eq:localbound_wtd},
\begin{equation}
  \bar\sigma(x)=2x\artanh(x) + 2x\sqrt{x \artanh(x)+\frac{x^2}{1-x^2}}.
  \label{eq:wtd_parametric}
\end{equation}

For the waiting time fluctuations of the flow, the bound coincides with the Fano factor above the critical entropy production $\bar\sigma_c$. Below, thanks to the linearity of $f_c(z)$ in Eq.~\eqref{eq:localbound_flow_wtd}, the optimisation can be performed analytically for given $\bar\sigma$ (with $\hat z_{1,2}$ determined numerically), such that no parametric form is needed.

The parametric forms of the bounds can be used to derive series expansions of the bounds for both small and large $\bar\sigma$. The results are given in Tab.~\ref{tab:asymptotics}. Thanks to convexity, the linear expansions for small $\bar\sigma$ are valid globally, but tight only for small $\bar\sigma$. The slope in the bound on flow fluctuations is the result of a transcendental equation (see Appendix~\ref{sec:smallsigma}) and has therefore been calculated numerically. The bounds for large $\bar\sigma$ are asymptotically exact and can be used to practically calculate the cost for high precision. They have been truncated at linear order in $F$ and $\varepsilon^2$, respectively. Already for a relative uncertainty of 10\,\% (i.e., $\varepsilon^2=0.01$ and analogously $F=0.01$), the difference between these approximations and the actual bounds is less than $0.05$, negligible compared to the two leading terms that give $\bar\sigma\gtrsim 209$.

\section{General quadratic form of the bound on the CGF for waiting time fluctuations}
\label{sec:hessian}

The expression to be maximised in Eq.~\eqref{eq:cgf_multiedge} can be expanded in quadratic order
as
$\boldsymbol{v}\cdot\boldsymbol{H}\boldsymbol{v}/2-z\boldsymbol{s}\cdot\boldsymbol{v}/\langle\dot
N\rangle$, where the vector
$\boldsymbol{v}$ has components $v_0=\xi-1$ and $v_a=\nu_a-1$ (starting the labelling of visible edges with $a=1$) and the vector
$\boldsymbol{s}$ has components $s_0=0$ and $s_a=c_a $. The Hessian matrix has the components
\begin{widetext}
\begin{equation}
  \boldsymbol{H}=
    \begin{pmatrix}
      -\bar\alpha & -c_1\beta_1&-c_2\beta_2&-c_3\beta_3&\cdots\\
      -c_1\beta_1 & -c_1(\gamma_1+1)+c_1^2 & c_1c_2 & c_1 c_3&\cdots\\
      -c_2\beta_2 & c_2c_1& -c_2(\gamma_2+1)+c_2^2 & c_2 c_3&\cdots\\
      -c_3\beta_3 & c_3c_1 & c_2 c_2 & -c_3(\gamma_3+1)+c_3^2 &\cdots\\
      \vdots &\vdots &\vdots &\vdots &\ddots
    \end{pmatrix} \, ,
\end{equation}
\end{widetext}
where $\bar\alpha \coloneqq \sum_a c_a\alpha_a$. The coefficients
$\alpha_a,\beta_a,\gamma_a$ depend on $x_a$ or $y_a$ like
Eqs.~\eqref{eq:coeff_wtd} or \eqref{eq:coeff_flow}, for
$a\in\mathcal{T}$ or $a\in\mathcal{F}$, respectively. The maximisation
of the quadratic form over $\boldsymbol{v}$ form then yields the optimal value $\boldsymbol{v}^*=z\boldsymbol{H}^{-1}\boldsymbol{s}/\dN$. It can be obtained through Gaussian elimination as
\begin{align}
    v^*_0&=-\frac{Yz/\dN}{(\bar\alpha-Z)(X-1)+Y^2}, \\
    v^*_a&=\frac{1}{\gamma_a+1}\frac{z/\dN}{X-1+\frac{Y^2}{\bar\alpha-Z}}-\frac{\beta_a}{\gamma_a+1}v^*_0,
\end{align}
with $X \coloneqq \sum_ac_a/(\gamma_a+1)$, $Y \coloneqq \sum_ac_a\beta_a/(\gamma_a+1)$, and $Z \coloneqq \sum_ac_a\beta_a^2/(\gamma_a+1)$.
We then obtain as bound on the CGF
\begin{equation}
    \mean{e^{z t^*}}\geq \frac{z}{\dN}-\frac{1}{2\dN^2}\boldsymbol{s}\cdot\boldsymbol{H}^{-1}\boldsymbol{s}z^2+\mathcal{O}(z^3)
\end{equation}
and
$\varepsilon^2\geq-\boldsymbol{s}\cdot\boldsymbol{H}^{-1}\boldsymbol{s}/2$ as bound on the uncertainty, which evaluates to the bound shown in Eq.~\eqref{eq:epsilon_C}.

\section{Refined bound for waiting time fluctuations of flow-type observables at small $\bar\sigma$}
\label{sec:smallsigma}

The bound~\eqref{eq:localbound_flow_wtd} of Sec.~\ref{sec:phasetrans} can be refined for small $\bar\sigma$. We start by splitting the function \eqref{eq:nonconvex} into $f(z)=g(z)+h(z)$ with
\begin{equation}
    g(z)=\frac{3z/2}{1+2z}\ln\frac{1-z}{1+2z}.
\end{equation}
and
\begin{equation}
    h(z)=\frac{6z^2}{1+2z}.
\end{equation}
The former gives the contribution of a visible edge to the overall scaled entropy production, which is constrained by $\bar\sigma\geq-2\sum_ac_ag(z_a)$. For a refined bound, we need to
minimise Eq.~\eqref{eq:localbound_flow_wtd_detailed} over the unknown values of $z_a$ and $c_a$ with this constraint
applied. The resulting bound $\varepsilon^2\geq C''/(1-C'')$ with
\begin{equation}
  C''=\min_z\left[\frac{1-z}{3}+\frac{z^2}{\bar\sigma/2+\max_{\{c_a,z_a\}|z,\bar\sigma}
      \sum_ac_a f(z_a)}\right]
  \label{eq:wtd_constr}
\end{equation}
is obtained in two steps: We first maximise in the denominator over $c_a$ and $z_a$ with the
constraint on $\bar\sigma$ and fixed $z=\sum_ac_az_a$, before minimising over
$z$. Without the constraint on $\bar\sigma$, the inner optimisation yields as maximum the convex hull
$f_c(z)$, attained for values of $z_a$ and $c_a$ for which
\begin{equation}
  \sum_ac_ag(z_a)=\frac{(\hat z_2-z)g(\hat z_1)+(z-\hat z_1)g(\hat z_2)}{\hat z_2-\hat z_1}.
\end{equation}
Plugging in the optimal value $z^*$ of Eq.~\eqref{eq:zopt}, we see a crossover
of this expression with $-\bar\sigma/2$ at $\bar\sigma_c'\simeq 1.5974$. For
smaller values of $\bar\sigma$, the maximum in Eq.~\eqref{eq:wtd_constr} must
be less than $f_c(z)$. Because of the uniqueness of the common tangent
defining the convex hull of $f(z)$, we do not expect any other local optima of
the unconstrained optimisation problem. Hence, the optimum of the constrained
problem must saturate the inequality, i.e., $\bar\sigma=-2\sum_ac_ag(z_a)$.
This constraint can now be accounted for through a Lagrange multiplier
$\lambda$, yielding
\begin{equation}
  \max_{\{c_a,z_a\}|z,\bar\sigma}\sum_ac_a f(z_a)=\max_{\{c_a,z_a\}|z}\sum_ac_a [f(z_a)+\lambda g(z_a)].
\end{equation}
This optimisation leads to another construction of a convex hull, as shown in
grey in Fig.~\ref{fig:convexhull}, in a convex region defined through the now
$\lambda$-dependent values $\hat z_{1,2}(\lambda)$, with $\hat z_{1,2}(0)=\hat
z_{1,2}$. The functions $\hat z_{1,2}(\lambda)$ need to be determined
numerically, and the Lagrange multiplier will need to satisfy
\begin{equation}
  \bar\sigma=\bar\sigma(z,\lambda) \coloneqq  -2\frac{(\hat z_2(\lambda)-z)g(\hat z_1(\lambda))+(z-\hat z_1(\lambda))g(\hat z_2(\lambda))}{\hat z_2(\lambda)-\hat z_1(\lambda)}.
\end{equation}
Then, the bound can be written as
\begin{equation}
  C''=\min_{z,\lambda|\bar\sigma}\left[\frac{1-z}{3}+\frac{z^2}{\hat h(\lambda,z)}\right]
\end{equation}
with
\begin{equation}
  \hat h(z,\lambda)=\frac{(\hat z_2(\lambda)-z)h(\hat z_1(\lambda))+(z-\hat z_1(\lambda))h(\hat z_2(\lambda))}{\hat z_2(\lambda)-\hat z_1(\lambda)}.
\end{equation}
Given the linearity of the constraint $\bar\sigma(z,\lambda)$ in $z$, we can
solve for $z$ and reduce the constrained optimisation to an unconstrained one
over the single parameter $\lambda$. The maximal value of $\lambda$ that needs
to be considered is $\lambda_\mathrm{max}\simeq 0.4569$, for which we obtain a
horizontal line at zero as common tangent to $f(z)+\lambda g(z)$, with
$\hat z_1(\lambda_\mathrm{max})=0$ and
$\hat z_2(\lambda_\mathrm{max})\simeq0.3772$. This is the only value of
$\lambda$ for which either $\hat z_1(\lambda)$ or $\hat z_2(\lambda)$ is zero,
and hence, given the convexity of $g(z)$ with $g(0)=0$, the only way of getting
$\bar\sigma(z,\lambda)=0$. Hence, $\bar\sigma=0$ requires $z=0$ and
$\lambda=\lambda_\mathrm{max}$, yielding $C''=1/3$ and $\varepsilon^2\geq
1/2$, as before for a single observed link in equilibrium.

\section{Mixed flow- and traffic-like counting observables}
\label{sec:mixed_obs}
The generic case is a counting observable involving a mix of flow
and traffic across the different edges. For a bound on the Fano factor, we follow the same steps as in Sec.~\ref{sec:multiedge_fano} to arrive at Eq.~\eqref{eq:fano_bound_general} with summation now over both $\mathcal{T}$ and $\mathcal{F}$. Crucially, the
coefficients $\alpha_a,\beta_a,\gamma_a$ take the functional form of
Eq.~\eqref{eq:coeff_traffic} for $a\in\mathcal{T}$ and of
Eq.~\eqref{eq:coeff_flow} for $a\in\mathcal{F}$.
With the substitutions (as before) $z_a=x_a^2$ for $a\in\mathcal{T}$ and
$z_a=y_a/(2-y_a)$ for $a\in\mathcal{F}$, we arrive at
\begin{align}
  F\geq &c_\mathrm{t}(1-z_\mathrm{t})+
  c_\mathrm{f}\frac{1}{2}(1-z_\mathrm{f})+\nonumber\\
  &\frac{(c_\mathrm{t}z_\mathrm{t}+c_\mathrm{f}z_\mathrm{f})^2}{\frac{\bar\sigma}{2}+c_\mathrm{t}(z_\mathrm{t}-\sqrt{z_\mathrm{t}}\artanh\sqrt{z_\mathrm{t}})+
  c_\mathrm{f}\frac{2z_\mathrm{f}(z_\mathrm{f}-\artanh z_\mathrm{f})}{1+z_\mathrm{f}}},
  \label{eq:fano_mixed}
\end{align}
where we have used Jensen's inequality individually for each of the partial
sums over $\mathcal{F}$ and $\mathcal{T}$ in the denominator. The partial
contributions to $\langle\dot N\rangle$ from $\mathcal{T}$ and $\mathcal{F}$
are $c_\mathrm{t} \coloneqq \sum_{a\in\mathcal{T}}c_a$ and
$c_\mathrm{f} \coloneqq \sum_{a\in\mathcal{F}}c_a$, respectively. Moreover, we have defined
$z_\mathrm{t} \coloneqq  \sum_{a\in\mathcal{T}}c_az_a/c_\mathrm{t}$ and $z_\mathrm{f} \coloneqq  \sum_{a\in\mathcal{F}}c_az_a/c_\mathrm{f}$.
Finally, we note that $z-\sqrt{z}\artanh\sqrt{z}\leq\frac{2z}{1+z}(z-\artanh
z)$ for $0\leq z<1$ (by visual inspection of the graphs and analysis of the asymptotics), giving us a weaker inequality with the same functional form for $z_\mathrm{t}$ as for
$z_\mathrm{f}$ in the denominator of Eq.~\eqref{eq:fano_mixed}. Finally, a Jensen
inequality for the two remaining terms yields Eq.~\eqref{eq:localbound_flow}
with $z \coloneqq  c_\mathrm{t}z_\mathrm{t}+c_\mathrm{f} z_\mathrm{f}$ and $y=2z/(1+z)$. Hence, the form of the bound on the precision
for pure flow observables, being weaker than the one for pure traffic
observables, holds also for a mix of both types. 

The bound derived in Sec.~\ref{sec:phasetrans} and Appendix~\ref{sec:smallsigma} for the waiting time
fluctuations for pure flow-type observables also holds for mixed flow-
and traffic-type observables. We start with
Eq.~\eqref{eq:wtd_general}, where the coefficients
$\alpha_a,\beta_a,\gamma_a$ take the form of Eq.~\eqref{eq:coeff_wtd}
for $a\in\mathcal{T}$ and of Eq.~\eqref{eq:coeff_flow} (applying to
both waiting time fluctuations and the Fano factor) for
$a\in\mathcal{F}$. Then, we substitute
$x_a=\sqrt{w_a}$ for $a\in\mathcal{T}$ and $y_a=2w_a/(w_a+1)$ for
$a\in\mathcal{F}$. For the first term in~\eqref{eq:wtd_general} we use
the inequality $2/(1-w)>(1-w)/(3-w)$, for the numerator of the second
term $w\geq 2w/(3-w)$, and for the denominator of the second term
\begin{equation}
  2w-\sqrt{w}\artanh\sqrt{w}\leq\frac{4w}{3-w}-\frac{2w}{1+w}\artanh(w)
\end{equation}
in the domain $0\leq w<1$, which can be verified by visual inspection
of the graphs and consideration of the asymptotic behaviour. Replacing
the form of the terms for the traffic by the form for the flow
therefore leads to a weaker bound. Re-substituting $y_a$ and then $z_a$, the bound takes
the form of Eq.~\eqref{eq:localbound_flow_wtd_detailed}, with the
summation now running over both $\mathcal{T}$ and
$\mathcal{F}$. Optimisation over $c_a$ and $z_a$ then leads to the
same bound as for pure flow-type observables.

\bibliography{trafficbound}
\end{document}